\documentclass[prl,letterpaper,english,reprint,nofootinbib,aps,superscriptaddress]{revtex4-2}

\usepackage{babel,calc,amsmath,amsthm,amssymb,graphicx,subfigure,comment}
\usepackage{mathdots}
\usepackage[T1]{fontenc}
\usepackage[unicode=true]{hyperref}
\usepackage{booktabs}
\usepackage{mathtools}
\usepackage{dsfont}
\usepackage{multirow}
\usepackage[normalem]{ulem}
\usepackage{multibib}

\newcommand{\nocontentsline}[3]{}
\newcommand{\tocless}[2]{\bgroup\let\addcontentsline=\nocontentsline#1{#2}\egroup}

\usepackage[dvipsnames]{xcolor}
\definecolor{prllinkblue}{rgb}{0.1804, 0.1882, 0.5726}
\hypersetup{
     colorlinks=true,       		
     linkcolor=prllinkblue,          	
     citecolor=prllinkblue,             
     urlcolor=prllinkblue,          
 }
 
\newtheorem*{theorem*}{Theorem}

\newtheorem*{corollary*}{Corollary}

\newtheorem*{lemma*}{Lemma}

\newtheorem*{proposition*}{Proposition}
\theoremstyle{definition}

\newtheorem*{definition*}{Definition}
\theoremstyle{remark}

\newtheorem*{remark*}{Remark}

\newcommand{\ket}[1]{\left\vert#1\right\rangle}
\newcommand{\bra}[1]{\left\langle#1\right\vert}
\newcommand{\braket}[1]{\left\langle#1\right\rangle}

\newcommand{\Tr}{{\rm Tr}}

\newif\ifdebug

\ifdebug
\definecolor{zhliu}{rgb}{0.12, 0.72, 0.36}
\definecolor{marker}{rgb}{0.96, 0.48, 0}
\newcommand{\add}[1]{\textcolor{zhliu}{#1}}
\newcommand{\note}[1]{\textcolor{marker}{#1}}
\newcommand\delete{\bgroup\markoverwith{\textcolor{marker}{\rule[0.5ex]{2pt}{0.8pt}}}\ULon}

\else
\newcommand{\add}[1]{#1}
\newcommand{\note}[1]{\ignorespaces}
\newcommand{\delete}[1]{\ignorespaces}

\fi

\begin{document}
\renewcommand{\figurename}{Fig.}


\title{Experimental Test of High-Dimensional Quantum Contextuality Based on Contextuality Concentration}


\author{Zheng-Hao~Liu}
\altaffiliation[Present address: ]{Center for Macroscopic Quantum States (bigQ), Department of Physics, Technical University of Denmark, Fysikvej, 2800 Kgs. Lyngby, Denmark}
\affiliation{CAS Key Laboratory of Quantum Information, University of Science and Technology of China, Hefei 230026, People's Republic of China}
\affiliation{CAS Centre For Excellence in Quantum Information and Quantum Physics, University of Science and Technology of China, Hefei 230026, People's Republic of China}

\author{Hui-Xian~Meng}
\affiliation{School of Mathematics and Physics, North China Electric Power University, Beijing 102206, People's Republic of China}
\affiliation{Theoretical Physics Division, Chern Institute of Mathematics, Nankai University, Tianjin 300071, People's Republic of China}

\author{Zhen-Peng Xu}
\affiliation{Naturwissenschaftlich-Technische Fakult\"{a}t, Universit\"{a}t Siegen, Walter-Flex-Stra{\ss}e 3, 57068 Siegen, Germany}

\author{Jie~Zhou}
\affiliation{Theoretical Physics Division, Chern Institute of Mathematics, Nankai University, Tianjin 300071, People's Republic of China}
\affiliation{Centre for Quantum Technologies, National University of Singapore 117543, Singapore}

\author{Jing-Ling~Chen}
\email{chenjl@nankai.edu.cn}
\affiliation{Theoretical Physics Division, Chern Institute of Mathematics, Nankai University, Tianjin 300071, People's Republic of China}

\author{Jin-Shi~Xu}
\email{jsxu@ustc.edu.cn}

\author{Chuan-Feng~Li}
\email{cfli@ustc.edu.cn}

\author{Guang-Can~Guo}
\affiliation{CAS Key Laboratory of Quantum Information, University of Science and Technology of China, Hefei 230026, People's Republic of China}
\affiliation{CAS Centre For Excellence in Quantum Information and Quantum Physics, University of Science and Technology of China, Hefei 230026, People's Republic of China}
\affiliation{Hefei National Laboratory, University of Science and Technology of China, Hefei 230088, People's Republic of China}

\author{Ad\'an~Cabello}
\email{adan@us.es}
\affiliation{Departamento de F\'{\i}sica Aplicada II, Universidad de Sevilla, E-41012 Sevilla,
Spain}
\affiliation{Instituto Carlos~I de F\'{\i}sica Te\'orica y Computacional, Universidad de
Sevilla, E-41012 Sevilla, Spain}


\date{\today}


\begin{abstract}
Contextuality is a distinctive feature of quantum theory and a fundamental resource for quantum computation. However, existing examples of contextuality in high-dimensional systems lack the necessary robustness required in experiments. Here we address this problem by identifying a family of noncontextuality inequalities whose maximum quantum violation grows with the dimension of the system. At first glance, this contextuality is the single-system version of multipartite Bell nonlocality taken to an extreme form. What is interesting is that the single-system version achieves the same degree of contextuality but uses a Hilbert space of {\em lower} dimension. That is, contextuality ``concentrates'' as the degree of contextuality per dimension increases. 
We show the practicality of this result by presenting an experimental test of contextuality in a seven-dimensional system. By simulating sequences of quantum ideal measurements with destructive measurements and repreparation in an all-optical setup,
we report a violation of $68.7$ standard deviations of the simplest case of the noncontextuality inequalities identified. Our results advance the investigation of high-dimensional contextuality, its connection to the Clifford algebra, and its role in quantum computation.
\end{abstract}


\maketitle


\textit{Introduction.---}In quantum theory, measurements cannot be considered as revealing preexisting properties that are independent of other compatible observables measured on the same system. This phenomenon is called contextuality or Kochen-Specker contextuality \cite{KS67,Budroni21}. It constitutes a fundamental resource for some quantum information processing tasks \cite{Cabello11, Amselem12} and some forms of universal quantum computation such as magic state distillation \cite{Bravyi05, Howard14} and measurement-based quantum computation \cite{Raussendorf13, Abramsky17, Frembs18}.

However, a fundamental problem is that arguably the most interesting forms of contextuality are experimentally inaccessible as they require high-dimensional quantum systems unavailable within current experimental platforms (for an extended discussion, see \cite{AC22}). This problem affects extreme forms of contextuality \cite{Cabello10b,Amaral15}, interesting temporal correlations \cite{Budroni:PRL2013,Budroni:PRL2014}, practical applications of contextuality such as dimension witnessing \cite{Guhne:2013PRA,Ray:2021NJP}, self-testing \cite{Kishor,Bharti:2019XXX}, sequential measurements-based machine learning \cite{Gao:2021XXX}, and topologically protected quantum computation \cite{LiuPRX21}. To attack this problem, one way is by looking for new high-dimensional systems \cite{AC22}. Another complementary approach is to identify forms of contextuality that are much more robust to noise.

The objective of this work is to produce robust contextuality in high-dimensional quantum systems.
The strategy we follow is looking for extreme forms of Bell nonlocality, which are multipartite versions of contextuality, and using the graph-theoretical approach to quantum correlations \cite{CSW14} to study single-particle versions of them. We observe that we can preserve the degree of contextuality but use a smaller dimensional quantum system. That is, there is a kind of ``concentration'' of contextuality in the transition between the multipartite and the single-particle cases. 
As a consequence, sequential measurements on a high-dimensional indivisible system can lead to quantum correlations whose violations of the corresponding noncontextuality inequalities grow with the system dimension. Moreover, the violations require Hilbert spaces smaller than that of the composite system manifesting the same degree of contextuality. This enhances the contextual correlation's robustness to noise and allows us to experimentally observe contextuality in high-dimensional systems. 
To demonstrate our findings, we report the experimental results of a path-encoded photonic qudit of $d=7$ which yields, when quantified by the quantum--classical ratio \cite{Amaral15}, the highest degree of contextuality ever observed on a single system.



\textit{Method.---}Bell nonlocality can be seen as a form of contextuality in which the requirement for compatibility is achieved using observables acting on spatially separated subsystems. Therefore, one can trivially convert every violation of a Bell inequality into a violation of a noncontextuality inequality that preserves both the degree of contextuality and the dimension of the Hilbert space \cite{Guehne10,Cabello21b}. 
A more intriguing approach is to associate every Bell operator with a graph indicating the exclusivity of a set of Bell experiment events \cite{CSW14}, that is, to specify which pairs of events are impossible to happen simultaneously. Then, by identifying a contextuality witness that shares the same graph of exclusivity, we can achieve a greater quantum violation and/or employ smaller dimensional systems \cite{Sadiq13,Rabelo14,Liu19,Ray21}.

Our starting point is the observation that the $n$-qubit Mermin-Ardehali-Belinskii-Klyshko (MABK) Bell inequalities \cite{Mermin90,Ardehali92,BK93} have maximum quantum violations that saturate the no-signaling bound and grow exponentially with the number of qubits. Using the graphs of exclusivity of each Bell MABK operator, we identify a family of noncontextuality inequalities that admit a single-particle realization. We then show that the minimal Hilbert space dimension required to achieve its maximal quantum violation is smaller than 
that needed to achieve the maximal quantum violation of the corresponding MABK inequalities. 
This phenomenon, hereafter called ``contextuality concentration'', is not limited to the MABK inequalities but also occurs for the 
bipartite three-settings 
Bell inequality \cite{CG04,Liu19} and, as shown here, for the Bell inequalities for graph states \cite{Guehne05,GC08,CGR08}. Our emphasis on the MABK inequalities is motivated by 
their high degree of nonlocality, resistance to noise, and low requirement of critical detection efficiency \cite{CRV08}. 


\textit{Extreme contextuality in high dimensions.}---The MABK inequalities for the $n$-party, two-setting, two-outcome or $(n,2,2)$ Bell scenarios with $n\ge 3$ odd 
\add{can be written as \cite{Mermin90}:
\begin{align}
{\cal M}_n &= \braket{M_n} \overset{\rm NCHV}{\leqslant} 2^{(n-1)/2},
\label{eq:MABKmean}
\end{align}
where ${\cal M}_n= \frac{1}{2 i} \sum_{\nu\in\{\pm 1\}} \nu\bigotimes_{j=1}^n \left(A_1^{(j)}+ i\nu A_2^{(j)}\right)$ and the operators $A_k^{(j)},\, k\in\{0,1,2\}$ have eigenvalues $\pm1$.
The superscripts differentiate the index of qubits, $\braket{\cdot}$ indicates expectation value and NCHV means the inequality holds for any noncontextual hidden-variable theory. Its maximal quantum violation, ${\cal M}_n = 2^{n-1}$, is achieved by choosing Pauli-like operators $\big[A_1^{(j)}, A_2^{(j)}\big] =2i A_0^{(j)}$ which pairwise anti-commute, and using the GHZ state $\ket{{\rm GHZ}_n} = \left( \bigotimes_{j=1}^n\ket{A_+^{(j)}} + i \bigotimes_{j=1}^n\ket{A_-^{(j)}} \right)/\sqrt{2}$, with $\ket{A_\pm^{(j)}}$ being the $\pm 1$-eigenstate of $A_0^{(j)}$.
}

To apply the graph-theoretic approach, \add{we rewrite $M_n$ as a linear combination of rank-1 projectors: $M_n=\sum_{k=1}^{2^{2n-2}} \Pi_k - \sum_{k=1}^{2^{2n-2}} \Pi_k^\prime$. The exponent $2n-2$ is due to that $M_n$ has $2^{n-1}$ terms and each term has $2^{n-1}$ positive (negative) projectors. By keeping only the projectors with positive signs, a witness of contextuality can be expressed in terms of event probabilities. Explicitly, $\mu_n = \sum_{k=1}^{2^{2n-2}}\braket{\Pi_k} = {\cal M}_n /2 + 2^{n-2}$.}
Let us call $G_n$ the graph of exclusivity of the events in $\mu_n$; 
we illustrate the case of $n=3$ in Fig.~\ref{fig:MABKexc} \add{and elaborate the procedure in Supplemental Material \cite{SM}.} 
According to the graph-theoretic approach, the 
noncontextual 
bound and quantum maximum of $\mu_n$ are
\begin{equation}
\mu_n \overset{\rm NCHV}{\leqslant} \alpha(G_n) = 2^{(n-3)/2}+2^{n-2} \overset{\rm Q}{\leqslant} \vartheta(G_n) = 2^{n-1},
\label{eq:MABKprob}
\end{equation}
where $\alpha(G_n)$ and $\vartheta(G_n)$ are the independence and Lov\'asz numbers of $G_n$, respectively \cite{CSW14}. 
The first observation is that the gap between noncontextuality and quantum theory is $\vartheta/\alpha=2-2\left/\middle(1+2^{(n-1)/2}\right)$, and thus increases with $n$. 


\begin{figure}[t]
    \centering
    \includegraphics[width=.96 \columnwidth]{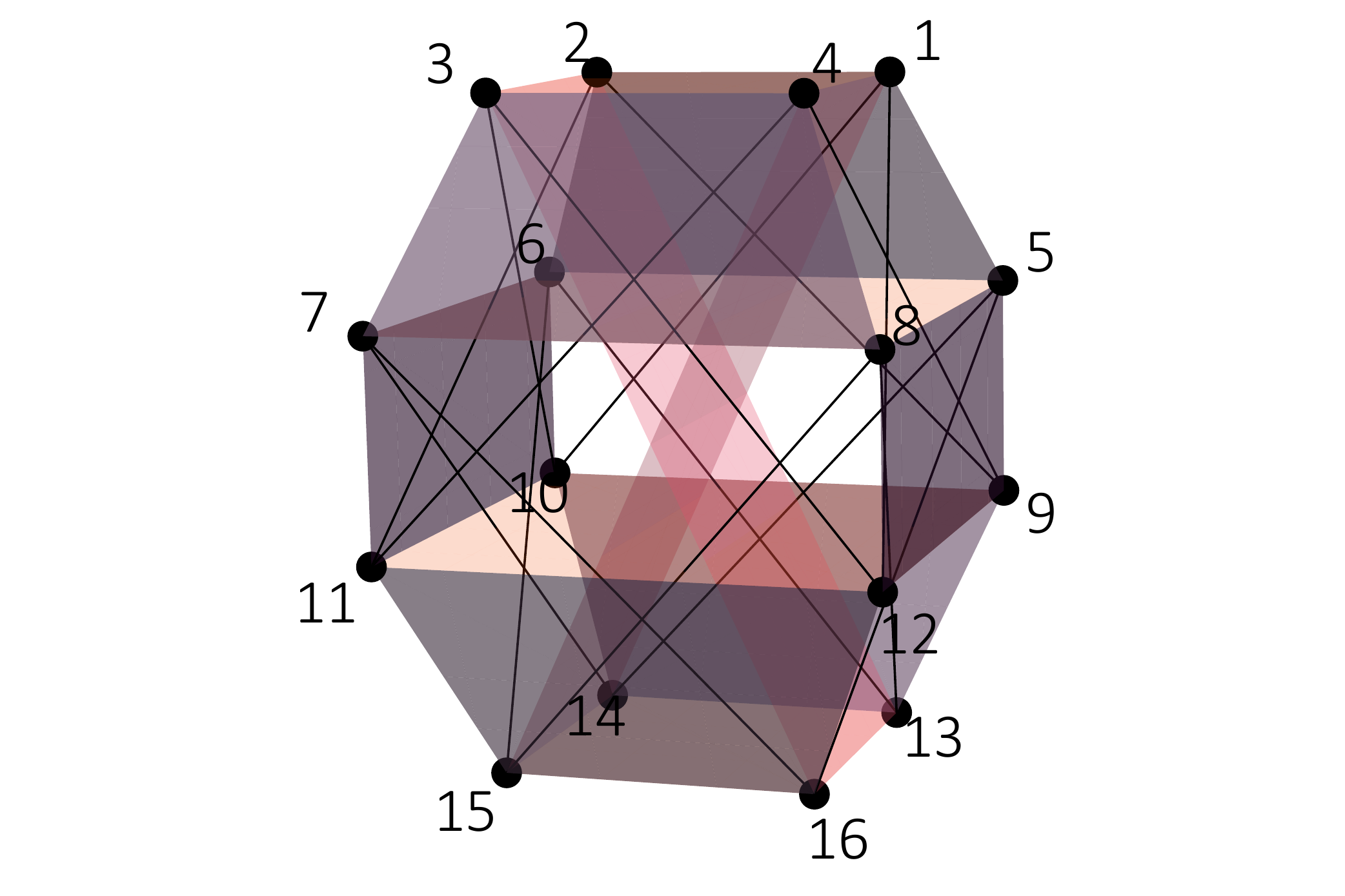}
    \caption{The graph of exclusivity associated with the events in $\mu_n$ for $n=3$.
    The points connected by a line represent pairs of mutually exclusive events. The four points on a colored quadrilateral represent four mutually exclusive events. 
    }
    \label{fig:MABKexc}
\end{figure}



We now proceed to show that the new graph-theoretic inequality (\ref{eq:MABKprob}) is stronger than the MABK inequality (\ref{eq:MABKmean}), in the sense that the quantum maximum of $\mu_n$ exploits only $2^n-1$-dimensional Hilbert space\add{---one less than in the $n$-qubit Bell scenario.} The proof is by explicit construction.
Let us denote 
the juxtaposition of the projectors in Eq.\,(\ref{eq:MABKprob}) as ${\cal A} = \left(\Pi_1\,\Pi_2\,\cdots\,\Pi_{2^{2n-2}} \right)$, then,
\begin{align}
    {\rm rank} ({\cal A}) = 2^n - {\rm dim (solution~space~of~}{\cal A}\mathbf{x} = \underbrace{\mathbf{00\cdots0}}_n),
\end{align}
where $\mathbf{x}$ is a $2^n$-dimensional ray. However, the only solution to ${\cal A}\mathbf{x} = 
\mathbf{00\ldots0}$ 
is a phase-flipped GHZ state: 
\add{
$A_0^{(1)}
\ket{{\rm GHZ}_n} = \left( \bigotimes_{j=1}^n\ket{A_+^{(j)}} - i \bigotimes_{j=1}^n\ket{A_-^{(j)}} \right)/\sqrt{2}. $
}To check the validity of the solution we observe that, as $A_0^{(1)}$ and $A_1^{(1)}$($A_2^{(1)}$) anti-commute, an additional $A_0^{(1)}$ in the state will cause every term in Eq.\,(\ref{eq:MABKmean}) to inverse sign; Therefore, for the phase-flipped GHZ state, ${\cal M}_n=-2^{n-1}$. Translating it into the event probabilities, we immediately find that $\mu_n$ evaluates to 0. The solution is unique because only the GHZ state maximally violates Eq.\,(\ref{eq:MABKmean}). Consequently, the projectors in ${\cal A}$ span only a $2^n-1$-dimensional space and can be realized in a quantum system with the same dimension. 

The above results show that some forms of multipartite nonlocality can be considered originating from contextuality in lower dimensions and,  reciprocally, that some forms of multipartite nonlocality can be ``concentrated'' into single-particle contextuality with a dimension advantage. In addition, the maximal quantum violation of Eq.\,(\ref{eq:MABKprob}) can be efficiently obtained by the semidefinite program of Lov\'asz optimization \cite{Ray:2021NJP}; one possible realization for the $n=3$ case is given in Supplementary Material \cite{SM}. This is in stark contrast with the situation in Bell nonlocality, where the maximal violation is not decidable even with a hierarchy of semidefinite programs \cite{Navascues15}.


\textit{High-dimensional contextuality without inequalities.}---Just as nonlocality can be revealed by Hardy- and GHZ-type proofs \cite{Hardy93,GHZ89} without using inequalities, the same can be done for contextuality \cite{Cabello13,Marques14,zpxu20}. However, no constructions of such proof are known for high-dimensional systems.

Here, we report a large class of logical contextuality from the exclusivity structures of the so-called graph states \cite{Hein04,note2}. 
Using Clifford algebra, we prove in Supplemental Material \cite{SM} that, if the representation of an $n$-qubit graph state has an odd number of vertices and at least one universal vertex, the events corresponding to the GHZ-type nonlocality produced by the graph state will induce a graph of exclusivity that can be implemented in a $2^n-1$-dimensional Hilbert space. 
Therefore, graph states are ideal candidates for showcasing examples of contextuality concentration. 
Moreover, our construction here can secure a $100\%$ success probability of observing the Hardy-like events violating noncontextuality, 
thus paving the way for a robust experimental observation of logical contextuality in high-dimensional systems.

We develop the case $n=3$, where the only graph state, up to local operations, is the GHZ state. In this case, the graph of exclusivity coincides with the one of the $n=3$ MABK inequality. Subjecting to the graph of exclusivity depicted in Fig.~\ref{fig:MABKexc}, the logical contextuality can be formulated as:
\begin{align}
    \textstyle \sum_{i=1}^4 P(1|i) = &1,~
    \textstyle \sum_{i=5}^8 P(1|i) = 1,~
    \textstyle \sum_{i=9}^{12} P(1|i) = 1, \nonumber \\[4pt]
    \midrule \midrule
    &\textstyle 1 \overset{\rm Q}= P_{\rm suc} := \textstyle \sum_{i=13}^{16} P(1|i) \overset{\rm NCHV}= 0.
    \label{eq:PHCP-Hardy}
\end{align}
Here, $P(1|i)$ denotes the probability that the measurement outcome of the observable $\Pi_i$ is 1 and \add{$P_{\rm suc}$ is the success probability for observing events forbidden in noncontextuality theories.} The proof of Eq.\,(\ref{eq:PHCP-Hardy}) and the settings of projectors achieving the quantum maximum are deferred to Supplemental Material \cite{SM}.


\textit{Experiment.}---We present an experimental test of the simplest case of the noncontextuality inequalities in \eqref{eq:MABKprob} with measurements on a seven-dimensional quantum system. The system is encoded in the photonic path degree of freedom and our experiment utilizes the techniques of spatial light modulation \cite{Lima18, xmhu20, Pierangeli19}.

The main technical challenge of the experiment is to acquire the statistics of two-point sequential measurements with a photonic seven-dimensional system, which is an open technical problem for dimensions high enough for observing contextuality concentration.
To this objective, \add{we have devised a quantum-inspired procedure to realize a non-demolition measurement}
with destructive measurements followed by a repreparation of the post-measurement state \cite{Cabello16}. The procedure allowed us to emulate a sequence of two ideal measurements (i.e., yielding the same outcome when repeated and not disturbing compatible observables) of two rank-one projectors $\Pi_i$ and $\Pi_j$ with simple prepare-and-measure experiments \cite{Canas16, yxiao17}. Concretely, it works as follows:
first, perform a destructive measurement of $\Pi_i$. If the measurement yields the outcome $1$, then prepare the state $|i\rangle$; if it yields the outcome $0$, then prepare $\ket{\psi}-\braket{i|\psi}\ket{i}$, that is, the initial state with the $+1$-eigenstate of $\Pi_i$ subtracted.
Finally, measure $\Pi_j$ on the reprepared state. \add{Crucially, the context-independent, repeatable, and minimally disturbing nature of the procedure would allow us to justify the assumption of noncontextuality by resorting to classical physics. Therefore, violation of a noncontextuality inequality via the procedure certifies the experiment itself is indeed manifesting contextuality.}

To witness the high-dimensional contextuality, we extracted three kinds of quantities from the results of the prepare-and-measure experiment: (i) the probabilities in Eq.\,(\ref{eq:PHCP-Hardy}) demonstrating the Hardy-type contextuality; (ii) the quality of exclusivity, i.e., ${\left\{ P(i,j|1,1) \mid [\Pi_i,\Pi_j]=0 \right\}}$ for establishing a lower bound of $\mu_3$ with realistic measurements and check the measurement repeatability; and (iii) the absence of signaling between compatible measurements for confirming the ideality of the measurement and showing the observed effect was indeed due to contextuality, instead of disturbance.
Explicitly, for estimating $\mu_3$ under imperfect exclusivity, we used the fact that \cite{Cabello16},
\begin{align}
    \mu_3\geqslant \sum_{k}P(1|k)-\sum_{(i,j)}P(1,1|i,j),
    \label{eq:MABK-exp}
\end{align}
where $k\in V(G_3)$ is an index associated with a vertex in $G_3$ and $(i,j) \in E(G_3)$ is an edge in $G_3$.
The no-signaling condition between pairs of compatible observables $\Pi_i$ and $\Pi_j$ can be verified by checking if all the following signaling factors vanish:
\begin{align}
        \varepsilon_{ij} &= P(1|i) - P(1, \_|i, j), ~~~
        \varepsilon_{ij}^\prime = P(1|i) - P(\_, 1|j, i), \nonumber \\
        \varepsilon_{ji} &= P(1|j) - P(1, \_|j, i), ~~~
        \varepsilon_{ji}^\prime = P(1|j) - P(\_, 1|i, j).
    \label{eq:Signaling}
\end{align}
That is, the marginal probability of one measurement is statistically independent of the other, regardless of the sequence of two measurements performed. Here, $P(1, \_|i, j) = P(1|j) P(1|i=1, j) + P(0|j) P(1|i=0, j)$ denotes the marginal probability of $\Pi_i$ yielding outcome $1$ when $\Pi_j$ is subsequently measured; similarly definitions hold for the other marginal probabilities.



\begin{figure}[t]
    \centering
    \includegraphics[width=.98 \columnwidth]{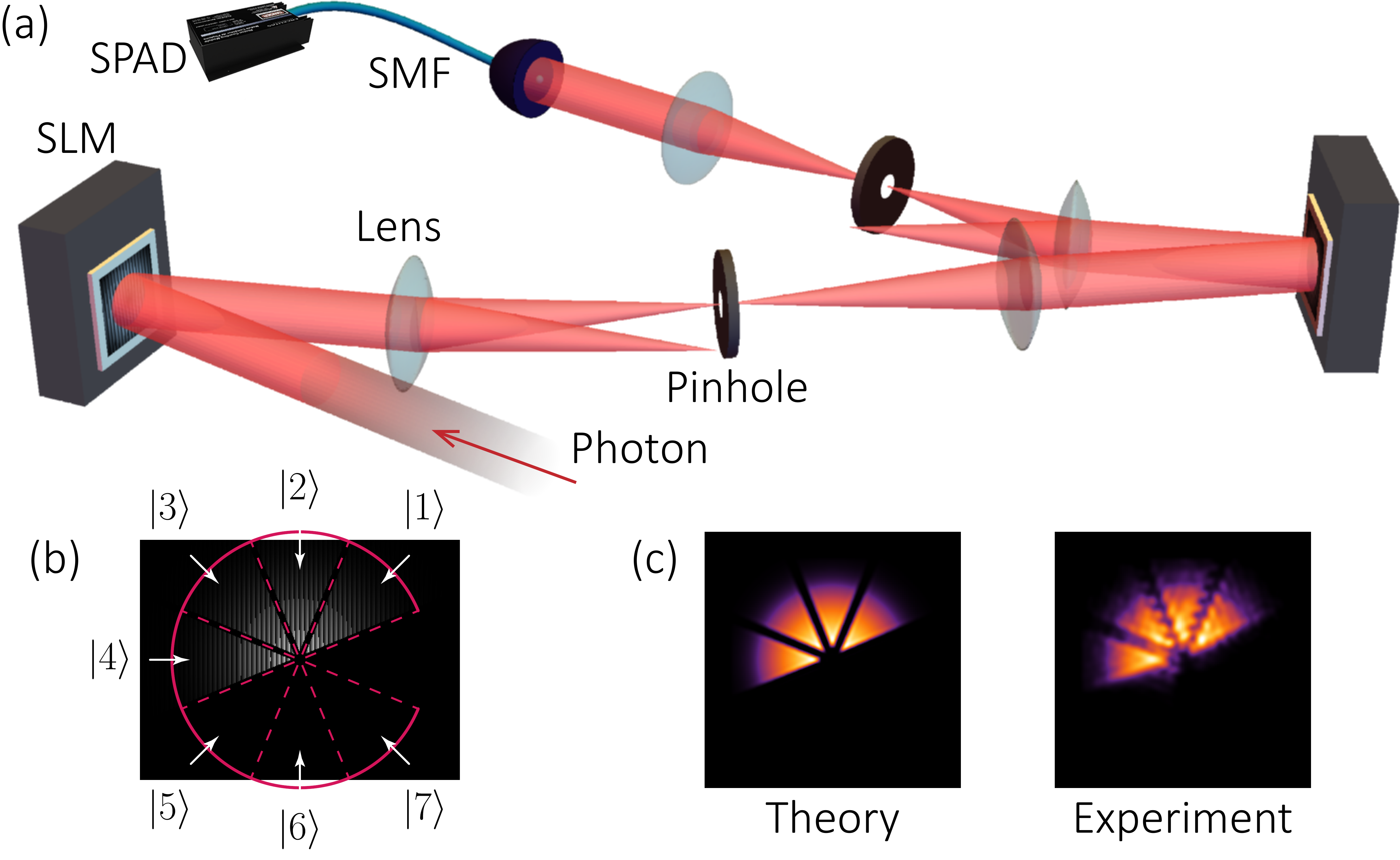}
    \caption{Experimental setup. 
    (a)~Optical setup of the prepare-and-measure experiment. A spatial light modulator (SLM)  encoded the initial state by modulating the photonic wavefront. A $4f$-correlator mapped the wavefront to the second SLM which implemented the measurement. The photons are collected by a simgle-mode fiber (SMF) and sent to the single-photon avalanche detector (SPAD) for photon counting.
    (b)~A sample hologram showing the encoding scheme. The corresponding state is $\ket{\psi}=(\ket{1}+\ket{2}+\ket{3}+\ket{4})/2$. 
    (c)~An example of calculated versus measured wavefunction profile. \add{The holograms and their characterization results for all states (projectors) needed in the experiment are shown in Supplemental Material \cite{SM}.}
    }
    \label{fig:setup}
\end{figure}



Our experimental setup is illustrated in Fig.~\ref{fig:setup}(a). Photons from an attenuated 800\,nm laser were expanded and the wavefront resembled a Gaussian beam with a waist radius of 1.6\,mm. Throughout the experiment, we used the spatial mode degree of freedom of the photons to register the seven-dimensional qudit, where the computational bases $\ket{1}$ through $\ket{7}$ are the angular states localized within a circular sector of the Gaussian beam.
To generate these angular states and their superpositions, we casted photons on a spatial light modulator (SLM) that displayed a phase-only hologram of seven circular sectors (cf. Fig.~\ref{fig:setup}(b)). The hologram in each sector displayed a blazed grating with a different phase range; consequently, a fraction of photons underwent diffraction, resulting in their propagation direction being altered towards the second SLM.
By adjusting the maximum phase variation of the grating to control the amplitude of photon wavefunctions in the seven sectors, we can realize the encoding of arbitrary qudit states \cite{Bolduc13}, with an explicit example shown in Fig.~\ref{fig:setup}(c).

The modulated photons then propagated through a $4f$-correlator, where the unwanted diffraction orders were filtered by a pinhole at the focal plane of the first lens. At the output plane of the correlator, a second SLM implemented the qudit measurement by employing the reverse transformation of the encoding process \cite{dAmbrosio13,Bent15}. In this way, choices of the initial states and the measurement settings were realized by displaying different holograms on the first and second SLM, respectively. 
To check the precision of the setup, we prepared all holograms used in the experiment, measured the spatial wavefunction profiles directly before the second SLM with a charge-coupled device camera, and compared them with theoretical predictions (cf. Supplemental Material \cite{SM}). The results revealed an average Pearson correlation \cite{corr} of 95.5\% between theoretical and experimental wavefunction profiles.
Finally, a telescope shrank the beam waist, and the photons were collected by a single-mode fiber (SMF) to determine the detection probability of an initial state on a specific measurement basis with photon counting.


\begin{figure}[t]
    \centering
    \includegraphics[width=.98 \columnwidth]{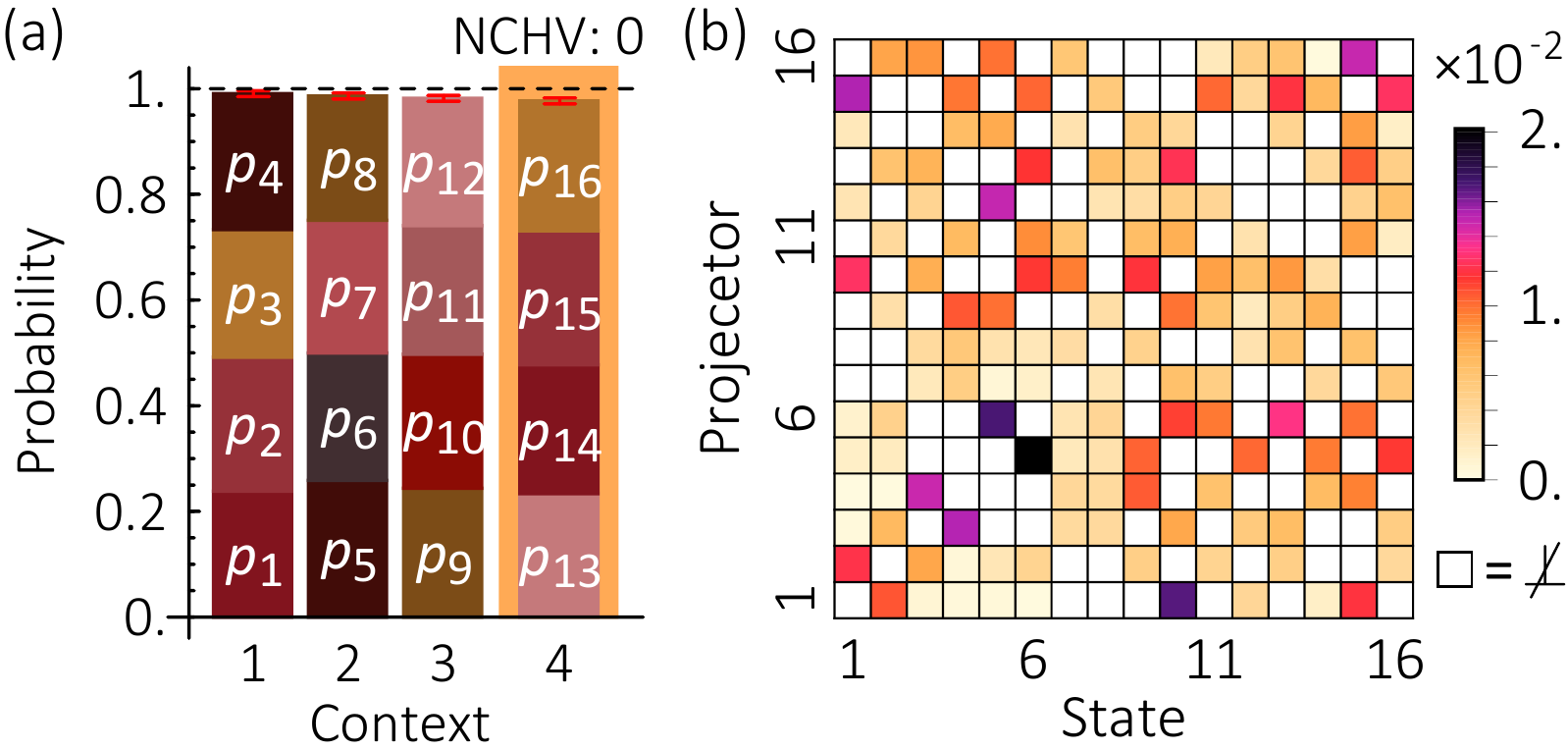}
    \caption{Experimental results. (a) Stacked bar chart of event probabilities with the 7-dimensional photonic system. Given all prerequisites in former columns. The last column with orange background has noncontextual (quantum) predictions of ideally 0(1), therefore manifesting logical contextuality. The error bars denote the $1\sigma$ standard deviations calculated by assuming a Poissonian counting statistics.
    (b)~Orthogonality between the projectors measured determined by preparing the nondegenerate eigenstate of every projector and measuring the detection probabilities on their corresponding compatible projectors. Only the grids corresponding to different compatible measurements are colored (the others are white).
    }
    \label{fig:result}
\end{figure}


Our experimental results are presented in Fig.~\ref{fig:result}. For the observation of Hardy-like contextuality (\ref{eq:PHCP-Hardy}), we displayed the hologram of the initial state on the first SLM and iterated the holograms on the second SLM over all the measurement basis. The total probabilities on the left-hand sides of the three constraints and the final nonclassical event were measured to exceed 97.6\%, exhibiting a sharp contradiction with the prediction from noncontextuality.
To ensure reasonable exclusivity of the compatible measurements in the experiment, pairs of holograms corresponding to orthogonal projectors were displayed on the two SLMs. The average detection probabilities for these settings were determined to be $\overline{P(1, 1|i, j)} = 0.64\%$---almost vanishing as expected for ideal measurements.
By substituting the recorded probabilities into Eq.\,(\ref{eq:MABK-exp}) to compensate for the deviations from ideal exclusivity and test the quantitative noncontextuality inequality, we found that $\mu_3 \geqslant 3.821 \pm 0.012$, violating the prediction of noncontextuality by 68.7 standard deviations. Here, the standard deviations were estimated by assuming a Poisson distribution for the statistics and resampling the recorded data (cf. Supplemental Material \cite{SM}).


\begin{figure}[t]
    \centering
    \includegraphics[width=.97 \columnwidth]{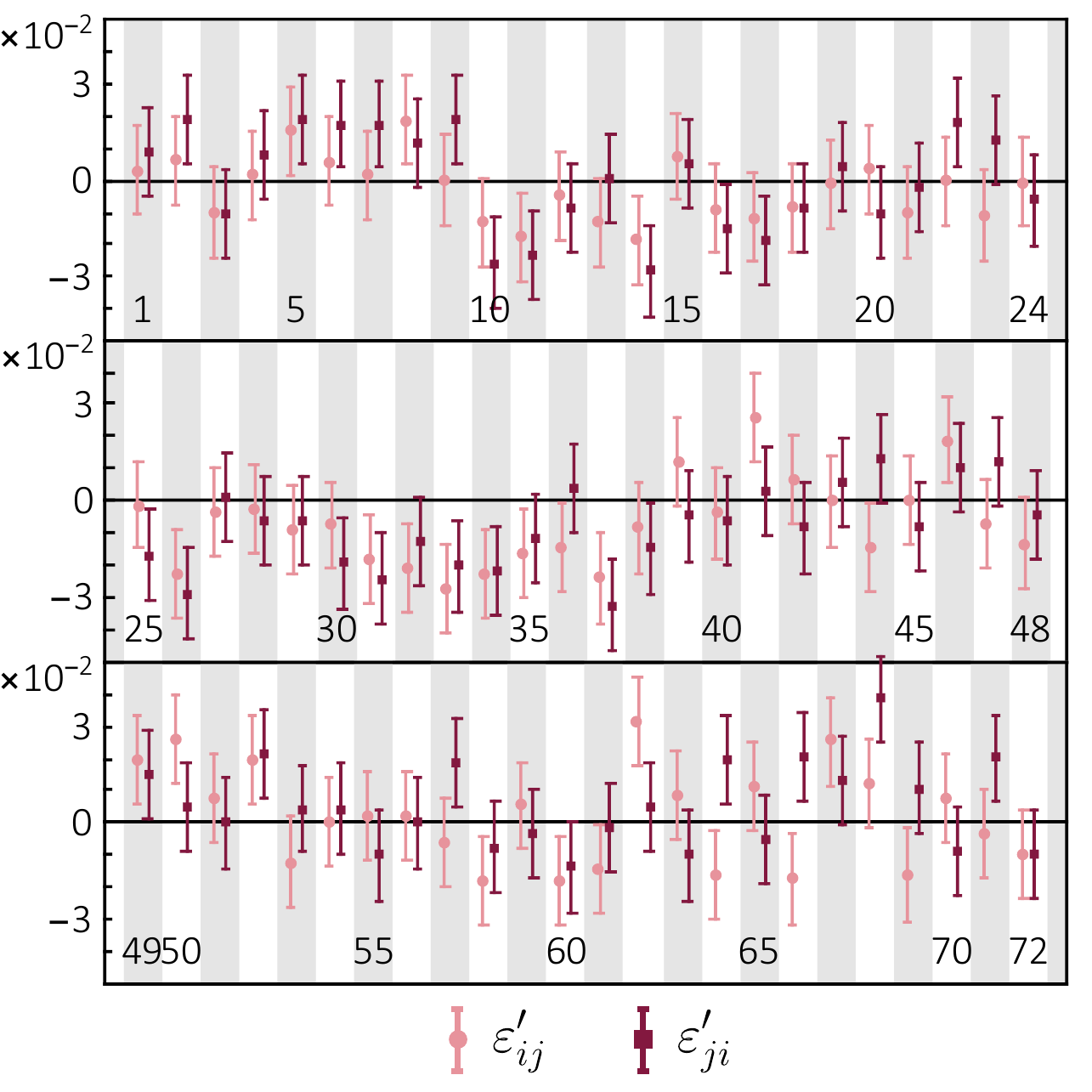}
    \caption{Verification of no-signaling condition. The numbers stand for indices of edges $(i, j) \in G_3$, sorted by $i$ then $j$ with $i<j$. The data points denote signaling factors and the error bars denote the $1\sigma$ standard deviations calculated by assuming a Poissonian counting statistics.
    }
    \label{fig:signal}
\end{figure}


To verify the no-signaling condition, 
we prepared the conditional state after the first projector measurement for each edge in the exclusivity graph $G_3$, and then measured it with the second projector. Owing to the normalization of counting, the effect of a later measurement upon an earlier one, represented by $\varepsilon_{ij}$ and $\varepsilon_{ji}$, would always be zero. The result of $\varepsilon_{ij}^\prime$ and $\varepsilon_{ji}^\prime$ in the no-signaling test for the 72 edges in $G_3$ is shown in Fig.~\ref{fig:signal}.
Most of the signaling factors are within one standard deviation from zero, with a quantitative characterization gave $\overline{|\varepsilon^\prime|} = (1.17 \pm 1.39)\%$. The small nonzero values can be considered to originate from experimental imprecision and small drifting over time, and the overall results complied well with the no-signaling requirement.


{\em Conclusions.---}We have identified quantum correlations resulting from sequential measurements on a single particle that manifest extremely strong forms of contextuality in lower dimensions. These correlations exhibit large violations of noncontextuality inequalities and perfect success probabilities for single-shot detection of contextuality.
An accompanying photonic experiment fleshed out the theoretical findings by simulating sequences of ideal quantum measurements with destructive measurements and observed the highest degree of contextuality on a single system (cf. Supplemental Material \cite{SM} for a comparison with previous studies). \add{Although the present result only shows a dimension reduction of 1, we envisage that contextuality concentration can be scalable by utilizing graph products, thereby offering additional advantages.} Because contextuality in stabilizer subtheory-based exclusivity structures investigated here forms the backbone of quantum computation architectures \cite{Bravyi05, Howard14, Raussendorf13, Abramsky17, Frembs18, Delfosse15, Shahandeh21, Booth21} in many different physical systems \cite{LiuPRX21, Souza11, Mamaev19, Egan21, hlhuang21, Waldherr14, Michaels21, Bourassa20, Abobeih21, Bourassa21, Asavanant19, Larsen19, Larsen21}, our work may stimulate the development of high-dimensional quantum information processing and novel quantum algorithms.



\tocless{
\begin{acknowledgments}
We would like to thank 
Yu~Meng and Ze-Yan~Hao for insightful comments and Shihao~Ru, Carles~Roch~i~Carceller, and Abhinav~Verma for helpful feedback on an earlier version of the paper.
This work was supported by
%
the Innovation Program for Quantum Science and Technology (Grant No.\ 2021ZD0301400),
the National Natural Science Foundation of China (Grants
Nos.\ 61725504,
11821404,
and U19A2075),
the Fundamental Research Funds for the Central Universities (Grant No.\ WK2030000056),  
and the Anhui Initiative in Quantum Information Technologies (Grant No.\ AHY060300).
J.L.C.\ was supported by the National Natural Science Foundations of China (Grant Nos.\ 12275136, 11875167 and 12075001), the 111 Project of B23045, and the Fundamental Research Funds for the Central Universities (Grant No.\ 3072022TS2503).
H.X.M.\ was supported by the National Natural Science Foundations of China (Grant No.\ 11901317).
Z.P.X.\ was supported by the Alexander Humboldt foundation.
A.C.\ is supported by the MCIN/AEI Project No.~PID2020-113738GB-I00, Project Qdisc (Project No.~US-15097, Universidad de Sevilla), with FEDER funds, and QuantERA grant SECRET (MCIN/AEI Project No.~PCI2019-111885-2).
This work was partially carried out at the USTC Center for Micro and Nanoscale Research and Fabrication.
Z.H.L.\ and H.X.M.\ contributed equally to this work.
\end{acknowledgments}



}

\onecolumngrid

\clearpage
\newpage
\setcounter{page}{1}
\appendix

\setcounter{equation}{0}
\setcounter{figure}{0}
\renewcommand{\theequation}{S\arabic{equation}}
\renewcommand{\thefigure}{S\arabic{figure}}

\begin{center}
{\large \bf Supplementary Material for ``Experimental Test of High-Dimensional Quantum Contextuality Based on Contextuality Concentration''}
\end{center}

\bigskip

In this Supplemental Material, 
we start by summarizing the graph-theoretical approach to contextuality, and use it to derive the simplest case of contextuality concentration from the tripartite Mermin--Ardehali--Belinskii--Klyshko (MABK) inequality to a noncontextuality inequality whose maximal quantum violation can be saturated using a set of 7-dimensional state and projectors. 
We then prove a construction of logical contextuality from a family of Greenberger--Horne--Zeilinger- (GHZ)-type paradoxes in graph states\,\cite{s-zhliu21php} and show the resulted logical contextuality can be achieved in a Hilbert space whose dimension is less than the graph state. 
After that, we showcase the simplest case among the above logical contextuality using its graph of exclusivity and give a set of 7-dimensional rays manifesting the logical contextuality with a success probability of 100\%. This setting also maximally violates the simplest case of the family of noncontextual hidden-variable (NCHV) inequalities in the main text and is used in our experimental test. 
Further, we present more details about the experiment and data analysis. 
Finally, we conclude with a summary of results in single-particle-based contextuality tests to show the merit of contextuality concentration for generating a high degree of contextuality. 

\tableofcontents

\section{A brief summary of the graph-theoretic approach to contextuality}

A modern method of bounding the set of behaviors of quantum correlation is the graph-theoretic approach\,\cite{s-CSW14}. We say a behavior is a set of probability distributions derived from projective measurements over arbitrary quantum states. Due to the orthogonality of measurements, some events cannot happen simultaneously. The graph of exclusivity $G$ essentially captures this impossibility: its vertices $V(G)$ represent the events of observing certain measurement outcomes, and its edges $E(G)$ connect pairs of exclusive events. The graph-theoretic approach treats correlations of general observables by expressing them in terms of linear combination of probabilities of   events.
	
Once a graph of exclusivity is given, the set of noncontextual and quantum correlations it can support is determined by the stable set polytope ${\sf STAB}(G)$ and theta body ${\sf TH}(G)$, respectively:
\begin{align}
& {\sf STAB}(G) := {\rm conv} \left\{ x\in\{0,1\}^{|V|} \middle\vert x_i x_j =0, \forall (i,j)\in E \right\}, \nonumber \\
& {\sf TH}(G) := \left\{ p\in\mathds{R}_+^{|V|} \middle\vert\, p_i=\left\vert \braket{\psi\vert v_i} \right\vert^2, \{\ket{v}\}~{\rm OR~of}~\bar{G} \right\},
\end{align}
where $\bar{G}$ is graph complement of $G$, $\rm conv$ and $\rm OR$ denote convex combination and orthogonal representation. The sum of event probabilities in NCHV and quantum theories are thus bounded by graph constants, namely, the independence number and Lov\'asz number \cite{s-Cabello16}:
\begin{align}
\sum_{i\in V(G)} P(1|i) - \sum_{(i,j)\in E(G)} P(1,1|i,j) \overset{\rm NCHV}{\leqslant} \alpha(G) \overset{\rm Q}{\leqslant} \vartheta(G).
\label{eq:CSW}
\end{align}
Here, $P(1,1|i,j)$ is the probability of observing (ideally) exclusive events $i, j$ simultaneously happen in one experiment; this term compensates for the deviation from exclusivity. Graphs with a large $\vartheta/\alpha$ ratio has the merit of producing significant inconsistency between noncontxtuality and quantum theories.

\section{From the MABK inequality to the graph of exclusivity}

The derivation in the main text, section ``\textit{Extreme contextuality in high dimensions}'' is based on the device-independent form of MABK inequality for better generality. However, contextuality concentration exists in more general scenarios and does not depend on knowledge about MABK inequality to be interpreted.  
In this section, we will reiterate the derivation in a pedagogical manner and using explicit Pauli operators. We hope that, by coming through the derivation here, even readers not familiar with multipartite Bell nonlocality and the graph-theoretical approach of contextuality should be able to follow the result and gain insight into contextuality concentration. In order to improve the readability, throughout this Supplemental Material, we will use the shorthand notation $\{X, Y, Z\} \equiv \{\sigma_x, \sigma_y, \sigma_z\} = \{ \bigg( \begin{matrix} 0 & 1 \\ 1 & 0 \end{matrix} \bigg), \bigg( \begin{matrix} 0 & -i \\ i & 0 \end{matrix} \bigg), \bigg( \begin{matrix} 1 & 0 \\ 0 & -1 \end{matrix} \bigg) \}$ to represent the Pauli operators, and $\{\Pi_{\pm x}, \Pi_{\pm y}, \Pi_{\pm z}\} \equiv \{\dfrac{\mathbb{I}+X}{2}, \dfrac{\mathbb{I}+Y}{2}, \dfrac{\mathbb{I}+Z}{2}\}$ to denote the projectors corresponding to their $\pm 1$-eigenstate. 

To keep the discussion self-contained, we rewrite the definition of the MABK inequality, which is the same as the Eq.\,(1) in the main text. For $n\geqslant3$ odd:
\begin{align}
    M_n = \frac{1}{2 i} & \left[ \bigotimes_{j=1}^n \left(A_1^{(j)}+iA_2^{(j)}\right) - \bigotimes_{j=1}^n \left(A_1^{(j)}-iA_2^{(j)}\right) \right] , \quad \quad {\cal M}_n = \braket{M_n} \overset{\rm NCHV}{\leqslant} 2^{(n-1)/2}.
    \label{eqs:MABKmean}
\end{align}
By explicit expansion, $M_n$ can also be expanded as:
\begin{align}
    M_n =& \sum_{k=0}^{(n-1)/2} (-1)^k \, {\boldsymbol\pi}\left(\bigotimes_{j=1}^{2k+1} A_2^{(j)} \bigotimes_{j=2k+2}^{n} A_1^{(j)}\right),
    \label{eq:MABKalt}
\end{align}
where the symbol ${\boldsymbol\pi}$ means the summation of permutations of the superscripts that give distinct products. The total number of terms in $M_n$ then can be calculated to be $\displaystyle\sum_{k=0}^{(n-1)/2} \bigg( \begin{matrix} n \\ 2k+1\end{matrix} \bigg) = 2^{n-1}$.

We now focus only on the simplest (3-qubit) case of the family of inequalities and use the Pauli operators to replace the abstract operators $A_k^{(j)},\, k\in\{0,1,2\}$. Although the choices of these operators are arbitrary, following Mermin's convention, we shall use $A_1^{(j)}=X^{(j)}, \ A_2^{(j)}=Y^{(j)},\ \text{and~} A_0^{(j)}=Z^{(j)}$ so the maximal quantum violation is achievable using the GHZ-state. Substituting the above convention into Eq.\,(\ref{eq:MABKalt}), we have:
\begin{align}
    M_3 =& {\boldsymbol\pi}\left(Y^{(1)} X^{(2)} X^{(3)} \right) - {\boldsymbol\pi}\left(Y^{(1)} Y^{(2)} Y^{(3)} \right)\nonumber\\
    &= Y^{(1)} X^{(2)} X^{(3)} +  X^{(1)} Y^{(2)} X^{(3)} + X^{(1)} X^{(2)} Y^{(3)} - Y^{(1)} Y^{(2)} Y^{(3)}, \quad\quad {\cal M}_3 = \braket{M_3} \overset{\rm NCHV}{\leqslant} 2.
    \label{eq:MABKmean3}
\end{align}
We can also check that the 3-qubit GHZ-state is $\ket{\rm GHZ_3} = (\ket{000} + i\ket{111})/\sqrt{2}$, and $\braket{{\rm GHZ_3} |M_3| {\rm GHZ_3}} = 4$.

To recast the MABK operator into event-probability form, we use the projector expansion $X = \Pi_{+x} - \Pi_{-x}$ and similar for $Y$. Substituting the projector expansion into Eq. (\ref{eq:MABKmean3}) yields:
\begin{align}
    M_3 = \big(& \Pi_{+y}^{(1)} - \Pi_{-y}^{(1)}\big) \otimes \big( \Pi_{+x}^{(2)} - \Pi_{-x}^{(2)}\big) \otimes \big( \Pi_{+x}^{(3)} - \Pi_{-x}^{(3)}\big) + \big( \Pi_{+x}^{(1)} - \Pi_{-x}^{(1)}\big) \otimes \big( \Pi_{+y}^{(2)} - \Pi_{-y}^{(2)}\big) \otimes \big( \Pi_{+x}^{(3)} - \Pi_{-x}^{(3)}\big) \nonumber\\
    &+ \big( \Pi_{+x}^{(1)} - \Pi_{-x}^{(1)}\big) \otimes \big( \Pi_{+x}^{(2)} - \Pi_{-x}^{(2)}\big) \otimes \big( \Pi_{+y}^{(3)} - \Pi_{-y}^{(3)}\big) - \big( \Pi_{+y}^{(1)} - \Pi_{-y}^{(1)}\big) \otimes \big( \Pi_{+y}^{(2)} - \Pi_{-y}^{(2)}\big) \otimes \big( \Pi_{+y}^{(3)} - \Pi_{-y}^{(3)}\big) \nonumber\\[7pt]
    = \Big(& \, 
    \color{Orange} \Pi_{-y}^{(1)} \Pi_{+y}^{(2)} \Pi_{+y}^{(3)} \,+\, \Pi_{-y}^{(1)} \Pi_{-y}^{(2)} \Pi_{-y}^{(3)} \,+\, \Pi_{+y}^{(1)} \Pi_{+y}^{(2)} \Pi_{-y}^{(3)} \,+\, \Pi_{+y}^{(1)} \Pi_{-y}^{(2)} \Pi_{+y}^{(3)} \nonumber\\  
    & \color{Orange} + \Pi_{+y}^{(1)} \Pi_{+x}^{(2)} \Pi_{+x}^{(3)} \,+\, \Pi_{+y}^{(1)} \Pi_{-x}^{(2)} \Pi_{-x}^{(3)} \,+\, \Pi_{-y}^{(1)} \Pi_{+x}^{(2)} \Pi_{-x}^{(3)} \,+\, \Pi_{-y}^{(1)} \Pi_{-x}^{(2)} \Pi_{+x}^{(3)} \nonumber\\  
    & \color{Orange} + \Pi_{-x}^{(1)} \Pi_{+y}^{(2)} \Pi_{-x}^{(3)} \,+\, \Pi_{-x}^{(1)} \Pi_{-y}^{(2)} \Pi_{+x}^{(3)} \,+\, \Pi_{+x}^{(1)} \Pi_{+y}^{(2)} \Pi_{+x}^{(3)} \,+\, \Pi_{+x}^{(1)} \Pi_{-y}^{(2)} \Pi_{-x}^{(3)} \nonumber\\  
    & \color{Orange} + \Pi_{+x}^{(1)} \Pi_{+x}^{(2)} \Pi_{+y}^{(3)} \,+\, \Pi_{+x}^{(1)} \Pi_{-x}^{(2)} \Pi_{-y}^{(3)} \,+\, \Pi_{-x}^{(1)} \Pi_{+x}^{(2)} \Pi_{-y}^{(3)} \,+\, \Pi_{-x}^{(1)} \Pi_{-x}^{(2)} \Pi_{+y}^{(3)} \color{black} \,\Big) 
    \label{eq:MABKexpand3}\\ 
    -\,& \Big( \, 
    \Pi_{+y}^{(1)} \Pi_{-y}^{(2)} \Pi_{-y}^{(3)} \,+\, \Pi_{+y}^{(1)} \Pi_{+y}^{(2)} \Pi_{+y}^{(3)} \,+\, \Pi_{-y}^{(1)} \Pi_{-y}^{(2)} \Pi_{+y}^{(3)} \,+\, \Pi_{-y}^{(1)} \Pi_{+y}^{(2)} \Pi_{-y}^{(3)} \nonumber\\  
    &+ \Pi_{-y}^{(1)} \Pi_{-x}^{(2)} \Pi_{-x}^{(3)} \,+\, \Pi_{-y}^{(1)} \Pi_{+x}^{(2)} \Pi_{+x}^{(3)} \,+\, \Pi_{+y}^{(1)} \Pi_{-x}^{(2)} \Pi_{+x}^{(3)} \,+\, \Pi_{+y}^{(1)} \Pi_{+x}^{(2)} \Pi_{-x}^{(3)} \nonumber\\  
    &+ \Pi_{+x}^{(1)} \Pi_{-y}^{(2)} \Pi_{+x}^{(3)} \,+\, \Pi_{+x}^{(1)} \Pi_{+y}^{(2)} \Pi_{-x}^{(3)} \,+\, \Pi_{-x}^{(1)} \Pi_{-y}^{(2)} \Pi_{-x}^{(3)} \,+\, \Pi_{-x}^{(1)} \Pi_{+y}^{(2)} \Pi_{+x}^{(3)} \nonumber\\  
    &+ \Pi_{-x}^{(1)} \Pi_{-x}^{(2)} \Pi_{-y}^{(3)} \,+\, \Pi_{-x}^{(1)} \Pi_{+x}^{(2)} \Pi_{+y}^{(3)} \,+\, \Pi_{+x}^{(1)} \Pi_{-x}^{(2)} \Pi_{+y}^{(3)} \,+\, \Pi_{+x}^{(1)} \Pi_{+x}^{(2)} \Pi_{-y}^{(3)} \,\Big). \nonumber
\end{align}
By comparing Eq.\,(\ref{eq:MABKexpand3}) with the definition of $\mu_n$ in the main text, we immediately find the composing projectors of $\mu_3$ are those projectors with an \textcolor{Orange}{orange} color. We shall define the sixteen projectors as $\Pi_1, \ldots, \Pi_{16}$, respectively, so that:
\begin{align}
    \mu_3 = \sum_{k=1}^{16} \braket{\Pi_k}.
\end{align} 
Note we have rearranged the projectors in the second equality of Eq.\,(\ref{eq:MABKexpand3}). The purpose of doing so is to let the graph of exclusivity have better symmetry.

We are now ready to extract the relation of exclusivity and the graph of exclusivity from the set of projectors. Two projectors $\Pi_i$ and $\Pi_j$ are exclusive if $\Pi_i\Pi_j=0$; for the projectors in the definition of $\mu_3$, they are the pairs containing $\Pi_{+x}$ and $\Pi_{-x}$ (or $\Pi_{+y}$ and $\Pi_{-y}$) acting on the same qubit, so they are bound not to give outcome $+1$ in the same round of experiment no matter what state is being tested. By applying this rule on the set of projectors, we find the pairs of indices $(i,j),\, i<j$ of exclusive projectors, and thus the edge set of the graph of exclusivity, $G_3$, to be:
\begin{align}
    V(G_3) = \{& (1, 2), (1, 3), (1, 4), (1, 5), (1, 6), (1, 10), (1, 12), (1, 14), (1, 15), (2, 3), (2, 4), (2, 5), (2, 6), (2, 9), \nonumber\\ & (2, 11), (2, 13), (2, 16), (3, 4), (3, 7), (3, 8), (3, 10), (3, 12), (3, 13), (3, 16), (4, 7), (4, 8), (4, 9), (4, 11), \nonumber\\ & (4, 14), (4, 15), (5, 6), (5, 7), (5, 8), (5, 9), (5, 12), (5, 14), (5, 16), (6, 7), (6, 8), (6, 10), (6, 11), \nonumber\\ & (6, 13), (6, 15), (7, 8), (7, 10), (7, 11), (7, 14), (7, 16), (8, 9), (8, 12), (8, 13), (8, 15), (9, 10), (9, 11), \nonumber\\ & (9, 12), (9, 13), (9, 14),  (10, 11), (10, 12), (10, 13), (10, 14), (11, 12), (11, 15), (11, 16), \nonumber\\ & (12, 15), (12, 16), (13, 14), (13, 15), (13, 16), (14, 15), (14, 16), (15, 16)\}
\end{align}
The edge set induces the graph of exclusivity, $G_3$ in Fig.\,1 in the main text. It is the graph complement of the Shrikhande graph. We also redraw the graph of exclusivity in Fig.\,\ref{fig:context-graph} in this Supplemental Material. By explicitly checking the graph constants of $G_3$, we found that $\alpha(G_3)=3$ and $\vartheta(G_3)=4$, thus recovering all affirmations in the Eq.\,(2) in the main text.

\newpage

\section{Logical contextuality from the graph states}

The graph states\,\cite{s-Hein04} form a family of highly entangled multipartite state, they can be represented using undirected, connected graphs. In the graph representation $\cal G$ of a graph state, every qubit is denoted by a vertex, and the connectivity of the graph determines the structure of entanglement. More specifically, let the adjacency matrix of the representation of the graph state be $C({\cal G})$, then the stabilizing operator for each of the qubits are defined as:
\begin{align}
    S^{(j)} \equiv X^{(j)} \prod_{k=1}^n (Z^{(k)})^{C_{jk}}.
    \label{eqs:stab}
\end{align}
Note that, following the convention of Reference \cite{s-Hein04}, we will state all the definitions and proofs using the Pauli operators, but it is also possible to establish the results in a device-independent manner like in the main text. A graph state $\ket{\cal G}$ is in turn defined as the common $+1$ eigenstate of all its stabilizing operators: $$S^{(j)}\ket{\cal G}=\ket{\cal G}, ~\forall~ j\in\{1,\cdots, n\}.$$ 

A GHZ-type paradox for nonlocality is a contradiction between local hidden-variable theory and the quantum theory upon the measurement result for some dichotomic $(\pm 1)$ observables, where their prediction about the product of these observables disagree with each other: $-1 {\rm ~vs.~} +1$. An observation of a GHZ-type paradox with the result in favor of quantum theory serves as a strong proof of quantum nonlocality. For every graph state with at least one universal vertex in its graph representation, a GHZ-type paradox can be formulated using the products of some stabilizing operators\,\cite{s-zhliu21php}. For self-containment, we briefly repeat the construction for an $n$-qubit graph state where $n$ is odd: let the universal vertex be labeled as 1, then a GHZ-type paradox can be formulated with the following operators:
\begin{align}
    S^{(1)},\;\;\;S^{(1)\theta(n)}S^{(2)},\;\;\;\left\{S^{(1)\theta(j)}S^{(j)}S^{(j+1)} \,\middle\vert\, j\in\{2,\ldots,n-1\} \right\},\;\;\;S^{(1)\theta(n)}S^{(n)}, 
    \label{eqs:graphGHZ}
\end{align}
with $\theta(j)=1+C_{n2}+\sum_{k=2,k\neq j}^{n-1} C_{k(k+1)}$. The product of these $m+1$ operators evaluates to $-1$ and $+1$ in local hidden-variable and quantum theories, therefore manifesting a GHZ-type paradox.

We now turn to investigate the exclusivity structure behind the measurements in (\ref{eqs:graphGHZ}). Notice that these observables can also be expressed using individual events, i.e., projective measurements. By normalization of probabilities, the projectors onto the $+1$ eigenstates of the observables suffices to demonstrate the GHZ-type paradox. Let us denote the set of $2^{n-1}$ projectors corresponding to the $k$-th observables in (\ref{eqs:graphGHZ}) by ${\cal E}_k = \{E_{k1}, E_{k2}, \cdots, E_{k(2^{n-1})}\}$, and the graph of exclusivity corresponds to the assemblage of events $\mathfrak{E}=\{{\cal E}_1, {\cal E}_2, \cdots, {\cal E}_{n+1}\}$ by $G$ (not to be confused with the calligraphic $\cal G$ denoting the graph state itself). The graph $G$ has an order of $|V(G)|=(n+1)2^{n-1}$, and the graph constants of interest are the independence number $\alpha(G)=n$ and the Lov\'asz number $\vartheta(G)=n+1$.

Our main goal in this section is to prove the exclusivity graph $G$ can be embedded in a $2^n-1$-dimensional Hilbert space, which nonetheless still supports a logical proof of contextuality with a success probability of 1 to trigger a set of events defying noncontextuality. To this aim, it suffices to prove the $+1$ eigenstates of the projectors in $\mathfrak{E}$ spans a Hilbert space with a dimension of only $2^n-1$, so it can be related to a set of $2^n-1$-dimensional vectors by a unitary isomorphism. The rest of proof exploits a result in Clifford algebra.

\begin{lemma*}
    Applying Pauli operator $\sigma_z$ on the $j$-th qubit of a graph state inverts only the sign of the eigenvalue of the stabilizing operator $S^{(j)}$, and leave the eigenvalues of other stabilizing operator invariant.
\end{lemma*}
\begin{proof}
    The expectation values of the stabilizing operators on the $\sigma_z$-modified graph state read $$\braket{S^{(j)}}_{Z^{(k)}\ket{\cal G}} = \Tr \left(S^{(j)} Z^{(k)} \ket{\cal G}\bra{\cal G} Z^{(k)}\right) = \Tr \left(Z^{(k)} S^{(j)} Z^{(k)} \ket{\cal G}\bra{\cal G}\right).$$ From the definition of the graph state and the stabilizing operator (\ref{eqs:stab}), direct calculation shows $$Z^{(k)} S^{(j)} Z^{(k)} = (1-2\delta_{jk})S^{(j)}.$$ Where $\delta_{jk}$ is the Kronecker-$\delta$. Substituting the last relation into the expectation values yields the desired proposition. 
\end{proof}

Because all stabilizing operators for a graph state commute, it is straightforward to check that modulating $\ket{\cal G}$ with $\sigma_z$ on different qubits can alternate the signs of the eigenvalues of all stabilizing operators one-by-one.

The above Lemma shows there exist a $\sigma_z$-modified graph state, 
\begin{align}
    \ket{\tilde{\cal G}} = \bigotimes_{j=1}^n \,(Z^{(j)})^{f(j)} \cdot \ket{\cal G},
\end{align}
where $f(j)\in\{0,1\}$, that is the common $-1$ eigenstate of all the operators in (\ref{eqs:graphGHZ}). This is because for all but the last operator a new stabilizing operator appears in every new term, so iterative application of the Lemma on every qubit can guarantee every of the first $n$ operators has an eigenvalue of $-1$ for the properly modified graph state via a Gaussian elimination over $f(j)$. Finally, observe every stabilizing operator appears in (\ref{eqs:graphGHZ}) even times, the last operator must evaluates to $-1$ in order to keep the product, that is the eigenvalue of an identity matrix, positive. 

Using the conversion between events probabilities and operator expectations, we arrive at $\Tr \left(E_{kl}\ket{\tilde{\cal G}}\bra{\tilde{\cal G}} \right)=0, \,\forall\, k\in\{1,\cdots,n+1\}, l\in\{1,\cdots,2^{n-1}\}$. As such the following equivalent statements hold: (1) the solution space for $\mathfrak{E}\mathbf{x}=\mathbf{0}$ has the dimension of 1, (2) the projectors $\mathfrak{E}$ has a rank deficiency, and (3) the graph of exclusivity $G$ supports vector realization in $2^n-1$-dimensions that manifests a logical noncontextuality paradox. This completes our proof for the logical contextuality originated from the graph states applicable in lower dimensional indivisible systems.

An important merit of the construction here is it secures a $100\%$ success probability of observing the events violating noncontextuality, as opposed to previous Hardy-type contextuality \cite{s-Cabello13, s-Marques14} and nonlocality proofs \cite{s-Hardy93, s-shjiang18, s-yhluo18, s-myang19Hardy}, where the success probabilities are less than unity and, even less than the fractional parts of the corresponding Bell and noncontextuality inequalities \cite{s-Mermin94, s-KCBS08, s-Lapkiewicz11, s-Sadiq13}, due to the setting for maximal violation of the inequalities fail to saturate probabilities of individual cliques. As stated in the main text, the ideal success probability paves the way for a robust experimental observation of logical contextuality in high-dimensional systems.

\section{Direct proof of the logical contextuality (3) in the main text}

\begin{figure}[htbp]
    \centering
    
\begin{minipage}{.44\textwidth}
    \begin{equation}
    \label{eqs:7dHardy}
    \begin{array}{ll}
        \displaystyle \sum_{i=1}^4p_i=1, &  \\
        \displaystyle  \sum_{i=5}^{8}p_i=1, &  \\
        \displaystyle  \sum_{i=9}^{12}p_i=1, &  
        \vspace{2pt}\\
   \hline
   \hline
\displaystyle \sum_{i=13}^{16}p_i\overset{\rm NCHV}=0. &
        \end{array}
    \end{equation}
\end{minipage}
\begin{minipage}{.55\textwidth}
    \includegraphics[width=.66 \textwidth]{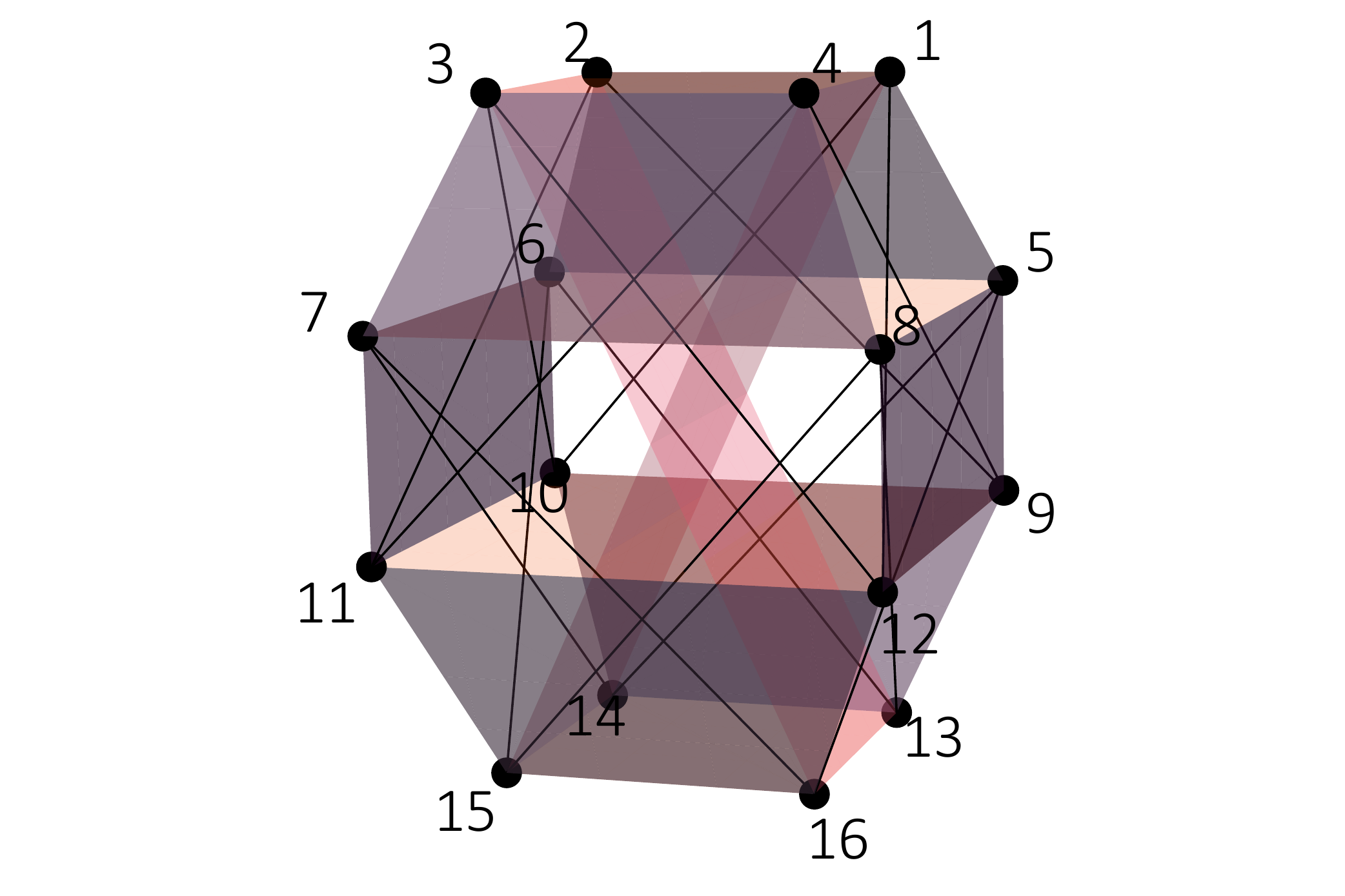}
\end{minipage}
    \caption{The 16 vertices represent 16 events. The points connected by a line denote exclusive events, and the four events represented by points on a colored quadrilateral are mutually exclusive.}
    \label{fig:context-graph}
\end{figure}

We show with the following example that the Hardy-type contextuality corresponding to the exclusivity graph in Fig.\,\ref{fig:context-graph} (same as the Fig. 1 in the main text) admits a logical proof. The proof can be constructed by saturation of subsets of exclusive probabilities; in a similar vein, the proof works for all Hardy-type contextuality with a $100\%$ success probability. Here in our case, the Hardy-like proof of contextuality can be explicitly stated by

\emph{Proof.}---The graph of exclusivity in Fig.\,\ref{fig:context-graph}is isomorphic to the strongly regular Shrikhande graph. Let us assume that the first three conditions in (\ref{eqs:7dHardy}) are satisfied. In the NCHV theory, the first condition in (\ref{eqs:7dHardy}) implies that there exists one and only one $p_i$ being $1$ and others being $0$, i.e., one and only one measurement direction in $\{| v_{i}\rangle:i=1,2,3,4\}$ with outcome $1$. In the following, we start the case study.

$(a)$ If $p_1=1$, then the outcome of measurement direction $| v_{1}\rangle$ is $1$.
The orthogonality relationships
$| v_1\rangle\bot | v_{5}\rangle,| v_{6}\rangle,$ $| v_{10}\rangle,| v_{12}\rangle,| v_{14}\rangle,| v_{15}\rangle$ result
that all outcomes of measurement directions $| v_{5}\rangle,| v_{6}\rangle,| v_{10}\rangle,| v_{12}\rangle,$ $| v_{14}\rangle,| v_{15}\rangle$ must be $0$, i.e., $p_{5}=p_{6}=p_{10}=p_{12}=p_{14}=p_{15}=0$. Thus, the second (third) condition in (\ref{eqs:7dHardy}) yields that  $p_{7}+p_{8}=1$ ($p_{9}+p_{11}=1$), i.e., the sum of the outcomes of  $| v_{7}\rangle$ and $| v_{8}\rangle$ (the sum of the outcomes of $| v_{9}\rangle$ and
$| v_{11}\rangle$) is $1$.

\ \ \ $(a_1)$ If the outcome of measurement direction $| v_7\rangle$ is $1$, then the outcome of
 $| v_{11}\rangle$ must be $0$ by  the orthogonality relationship $| v_7\rangle\bot | v_{11}\rangle$, and so $| v_{9}\rangle$'s outcome must be $1$ by the third condition $p_{9}+p_{11}=1$.
Since $| v_{13}\rangle\bot | v_{9}\rangle$, $| v_{14}\rangle\bot | v_{1}\rangle$, $| v_{15}\rangle\bot | v_{1}\rangle$, $| v_{16}\rangle\bot | v_{7}\rangle$, and all outcomes of
$| v_{1}\rangle,| v_{7}\rangle,| v_{9}\rangle$ are $1$, we obtain that the outcome of $| v_i\rangle$ ($i=13,14,15,16$) is $0$, i.e., $p_{13}+p_{14}+p_{15}+p_{16}=0$.

\ \ \ $(a_2)$ If the outcome of measurement direction $| v_8\rangle$ is $1$, then the outcome of
 $| v_{9}\rangle$ must be $0$ by  the orthogonality relationship $| v_8\rangle\bot | v_{9}\rangle$, and so $| v_{11}\rangle$'s outcome must be $1$ by the third condition $p_{9}+p_{11}=1$.
Since $| v_{13}\rangle\bot | v_{8}\rangle$, $| v_{14}\rangle\bot | v_{1}\rangle$, $| v_{15}\rangle\bot | v_{1}\rangle$, $| v_{16}\rangle\bot | v_{11}\rangle$, and all outcomes of
$| v_{1}\rangle,| v_{8}\rangle,| v_{11}\rangle$ are $1$, we obtain that the outcome of $| v_i\rangle$ ($i=13,14,15,16$) is $0$, i.e., $p_{13}+p_{14}+p_{15}+p_{16}=0$.

Due the the vertex transitivity of the graph, for the other cases: (b) $p_2=1$, (c) $p_3=1$, and (d) $p_4=1$, we can obtain $p_{13}+p_{14}+p_{15}+p_{16}=0$.
Therefore, in the NCHV theory, the first three conditions in (\ref{eqs:7dHardy}) lead to $p_{13}+p_{14}+p_{15}+p_{16}=0$. \hfill $\Box$


\section{Experimental details}

\paragraph{Analytical settings of projectors.---}\!\!\!\!Our experimental settings for realizing the maximal quantum violation of noncontextuality inequality (2) and observing the logical contextuality (3) in the main text are summarized in Table~\ref{tab:settings}. Here, $\ket{\psi}$ is the initial state and $\ket{v_i}$ represent the $+1$-eigenstate of the $i$-th projector. In other words, the projectors used in the experiment and the rays in the table are related via $\Pi_i=\ket{v_i}\bra{v_i}$. 
The other state preparations and measurement rays required in the experiment were all calculated from the rays listed in the table. Concretely, the full sets of compatible measurements from every four projectors were constructed by introducing extra projectors according to the symmetric form of the existing projectors, and when the projector $\Pi_i$ giving outcome 0 the post-measurement state was re-prepared as $(\mathbb{I}-\Pi_i)\ket{\psi}$ according to the L\"uders rule. 

\begin{table}[htbp]
    \caption{Optimal measurement settings $\ket{v_i}$ and input state $\ket{\psi}$ for realizing the maximal quantum violation of noncontextuality inequality (2) and observing the logical contextuality (3) in the main text in the 7-dimensional space. \\}
    \label{tab:settings}
    \centering
    \begin{tabular}{ccccccccccccccccc}
    \toprule
        $\ket{v_i}$ & $i_1$ & $i_2$ & $i_3$ & $i_4$ & $i_5$ & $i_6$ & $i_7$ & \phantom{123123123123} & $\ket{v_i}$ & $i_1$ & $i_2$ & $i_3$ & $i_4$ & $i_5$ & $i_6$ & $i_7$ \\ \midrule
        $\ket{v_1}$ & 1 & 0 & 0 & 0 & 0 & 0 & 0 & & $\ket{v_9}$ & 1/2 & 0 & 1/2 & 0 & 1/$\sqrt{8}$ & -1/$\sqrt{8}$ & 1/2 \\
        $\ket{v_2}$ & 0 & 1 & 0 & 0 & 0 & 0 & 0 & & $\ket{v_{10}}$ & 0 & 1/2 & 0 & 1/2 & -1/$\sqrt{8}$ & 1/$\sqrt{8}$ & 1/2 \\
        $\ket{v_3}$ & 0 & 0 & 1 & 0 & 0 & 0 & 0 & & $\ket{v_{11}}$ & 1/2 & 0 & 1/2 & 0 & -1/$\sqrt{8}$ & 1/$\sqrt{8}$ & -1/2\\
        $\ket{v_4}$ & 0 & 0 & 0 & 1 & 0 & 0 & 0 & & $\ket{v_{12}}$ & 0 & 1/2 & 0 & 1/2 & 1/$\sqrt{8}$ & -1/$\sqrt{8}$ & -1/2\\
        $\ket{v_5}$ & 0 & 0 & 1/2 & 1/2 & 0 & 1/$\sqrt{2}$ & 0 & & $\ket{v_{13}}$ & 1/2 & 0 & 0 & 1/2 & 1/$\sqrt{8}$ & 1/$\sqrt{8}$ & -1/2\\
        $\ket{v_6}$ & 0 & 0 & 1/2 & 1/2 & 0 & -1/$\sqrt{2}$ & 0 & & $\ket{v_{14}}$ & 0 & 1/2 & 1/2 & 0 & -1/$\sqrt{8}$ & -1/$\sqrt{8}$ & -1/2 \\
        $\ket{v_7}$ & 1/2 & 1/2 & 0 & 0 & 1/$\sqrt{2}$ & 0 & 0 & & $\ket{v_{15}}$ & 0 & 1/2 & 1/2 & 0 & 1/$\sqrt{8}$ & 1/$\sqrt{8}$ & 1/2 \\
        $\ket{v_8}$ & 1/2 & 1/2 & 0 & 0 & -1/$\sqrt{2}$ & 0 & 0 & & $\ket{v_{16}}$ & 1/2 & 0 & 0 & 1/2 & -1/$\sqrt{8}$ & -1/$\sqrt{8}$ & 1/2 \\
        \midrule
        &&&&&&&&& $\ket{\psi}$ & 1/2 & 1/2 & 1/2 & 1/2 & 0 & 0 & 0 \\
    \bottomrule
    \end{tabular}
  \end{table}

\bigskip

\paragraph{Estimation of counting errors in the experiment.---}The data in our experiment were registered with a single-photon avalanche detector, and the counting rate was about $10^5$/s when the preparation and measurement basis were aligned. As the counting events occurred randomly in time, but were independent of each other and identically distributed, the statistics of photon counting must follow a Poissonian distribution. 
To estimate the uncertainty of our data collection process, we took the registered number of counting events as the expectation of the Poissonian distribution, and resampled the numbers in a \texttt{Mathematica} program. We used the resampled data to generate 100 groups of detection event probabilities, which were further used to calculate the expectations of all the observables required in the experiment. The standard deviations of these expectations were taken as the $1\sigma$ error bars of the corresponding observables.

\vfill

\paragraph{Calibration of preparations and measurements.---}\!\!\!\!To benchmark the performance of our prepare-and-measure setup, we prepared the initial state and the $+1$-eigenstates of all the projectors that appeared in the experiment using the first and the second spatial light modulator (SLM), respectively, according to the analytical settings in Table~\ref{tab:settings}, and measured the beam profile at the focal plane conjugate to the second SLM using a charge-coupled device camera. Here, the projectors' $+1$-eigenstates were used to calibrate the performance of the measurements implemented by the projectors. In both cases, the other SLM displayed a blazed grating with maximal contrast to keep the optical path unchanged and maximize the optical power at the camera.
In Table~\ref{tab:imgcorr}, we report the Pearson correlations between the beam profiles corresponding to the state/projectors measured and the ideal intensity distribution.

\begin{table}[b!]
    \centering
    \caption{Experimentally prepared beam profiles, theoretical predictions, and the Pearson correlations between pairs of images. \\}
    \begin{tabular}{ccccccc}
        \toprule
        Ray & $\ket{v_1}$ & $\ket{v_2}$ & $\ket{v_3}$ & $\ket{v_4}$ & $\ket{v_5}$ & $\ket{v_6}$ \\
        \midrule
        Theory & 
        \raisebox{-.45\height}{\includegraphics[width=60pt]{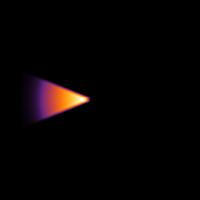}} & 
        \raisebox{-.45\height}{\includegraphics[width=60pt]{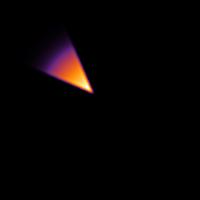}} & 
        \raisebox{-.45\height}{\includegraphics[width=60pt]{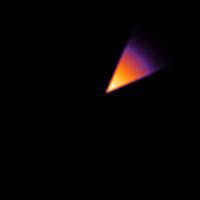}} & 
        \raisebox{-.45\height}{\includegraphics[width=60pt]{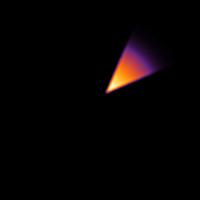}} & 
        \raisebox{-.45\height}{\includegraphics[width=60pt]{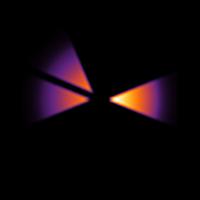}} & 
        \raisebox{-.45\height}{\includegraphics[width=60pt]{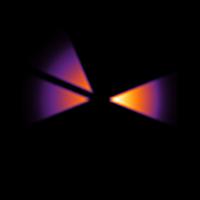}} \\[27pt]
        Experiment & 
        \raisebox{-.45\height}{\includegraphics[width=60pt]{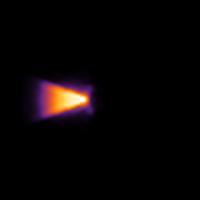}} & 
        \raisebox{-.45\height}{\includegraphics[width=60pt]{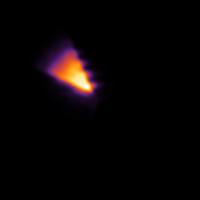}} & 
        \raisebox{-.45\height}{\includegraphics[width=60pt]{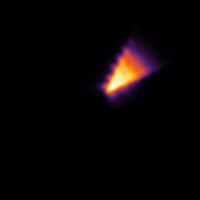}} & 
        \raisebox{-.45\height}{\includegraphics[width=60pt]{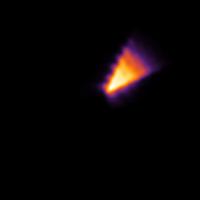}} & 
        \raisebox{-.45\height}{\includegraphics[width=60pt]{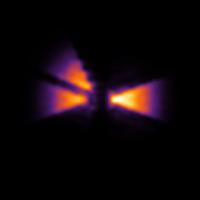}} & 
        \raisebox{-.45\height}{\includegraphics[width=60pt]{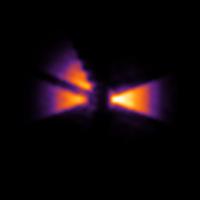}} \\
        \raisebox{.2\height}{\parbox[l]{5em}{Pearson\\correlation}} & 95.96\% & 95.86\% & 95.75\% & 95.75\% & 94.71\% & 94.79\% \\
        \multicolumn{7}{r}{Continued on next page $\to$} \\
        \bottomrule
    \end{tabular}
    \label{tab:imgcorr}
\end{table}

\setcounter{table}{1}

\begin{table}[htbp]
    \centering
    \caption{(continue)}
    \vspace{10pt}
    \begin{tabular}{ccccccc}
        \toprule
        Vector & $\ket{v_7}$ & $\ket{v_8}$ & $\ket{v_9}$ & $\ket{v_{10}}$ & $\ket{v_{11}}$ & $\ket{v_{12}}$ \\
        \midrule
        Theory & 
        \raisebox{-.45\height}{\includegraphics[width=60pt]{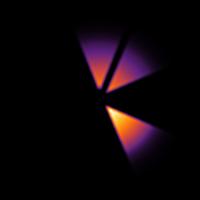}} & 
        \raisebox{-.45\height}{\includegraphics[width=60pt]{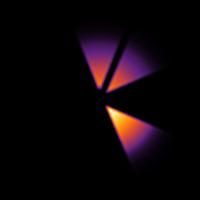}} & 
        \raisebox{-.45\height}{\includegraphics[width=60pt]{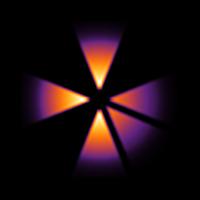}} & 
        \raisebox{-.45\height}{\includegraphics[width=60pt]{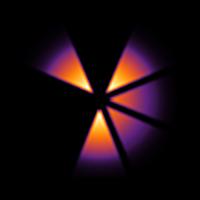}} & 
        \raisebox{-.45\height}{\includegraphics[width=60pt]{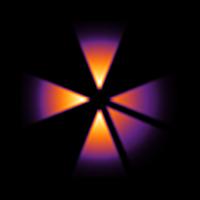}} & 
        \raisebox{-.45\height}{\includegraphics[width=60pt]{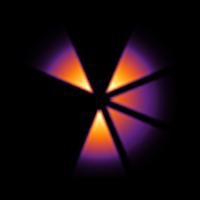}} \\[27pt]
        Experiment & 
        \raisebox{-.45\height}{\includegraphics[width=60pt]{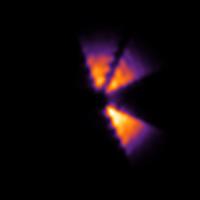}} & 
        \raisebox{-.45\height}{\includegraphics[width=60pt]{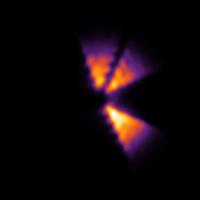}} & 
        \raisebox{-.45\height}{\includegraphics[width=60pt]{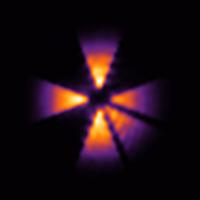}} & 
        \raisebox{-.45\height}{\includegraphics[width=60pt]{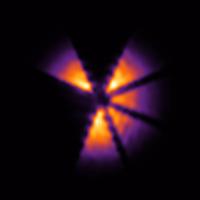}} & 
        \raisebox{-.45\height}{\includegraphics[width=60pt]{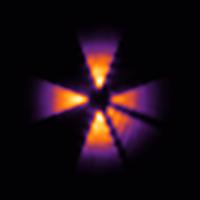}} & 
        \raisebox{-.45\height}{\includegraphics[width=60pt]{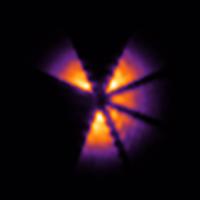}} \\
        \raisebox{.2\height}{\parbox[l]{5em}{Pearson\\correlation}} & 95.28\% & 95.24\% & 95.53\% & 95.07\% & 94.99\% & 94.67\% \\
        \midrule
        Vector & $\ket{v_{13}}$ & $\ket{v_{14}}$ & $\ket{v_{15}}$ & $\ket{v_{16}}$ & & $\ket{\psi}$ \\
        \midrule
        Theory & 
        \raisebox{-.45\height}{\includegraphics[width=60pt]{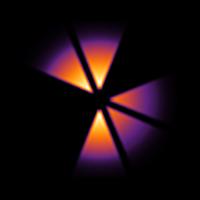}} & 
        \raisebox{-.45\height}{\includegraphics[width=60pt]{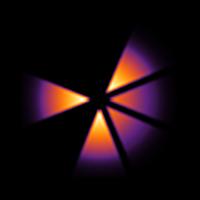}} & 
        \raisebox{-.45\height}{\includegraphics[width=60pt]{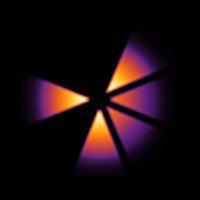}} & 
        \raisebox{-.45\height}{\includegraphics[width=60pt]{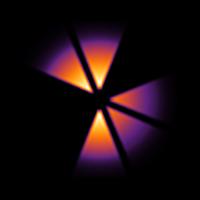}} & & 
        \raisebox{-.45\height}{\includegraphics[width=60pt]{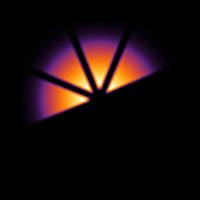}} \\[27pt]
        Experiment & 
        \raisebox{-.45\height}{\includegraphics[width=60pt]{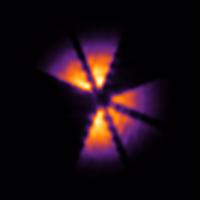}} & 
        \raisebox{-.45\height}{\includegraphics[width=60pt]{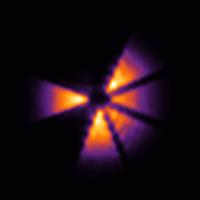}} & 
        \raisebox{-.45\height}{\includegraphics[width=60pt]{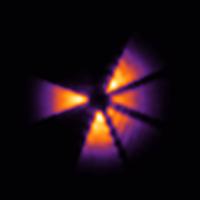}} & 
        \raisebox{-.45\height}{\includegraphics[width=60pt]{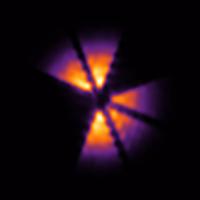}} & & 
        \raisebox{-.45\height}{\includegraphics[width=60pt]{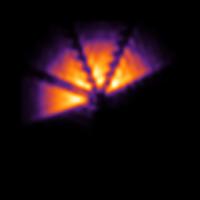}} \\
        \raisebox{.2\height}{\parbox[l]{5em}{Pearson\\correlation}} & 95.16\% & 95.31\% & 95.91\% & 95.64\% & & 95.82\% \\
        \midrule 
        &&&&\multicolumn{2}{c}{Average Pearson correlation} & 95.46\% \\
        \bottomrule
    \end{tabular}
\end{table}

\section{Data analysis}

\paragraph{Observation of the Hardy-type contextuality.---}We begin the data analysis by observing the Hardy-type contextuality in Eq.\,(3) of the main text. To this aim, only the detection probabilities of the initial state on the $+1$-eigenstate of the 16 projectors in Table~\ref{tab:settings} are required; these values, together with the sum of the probabilities corresponding to the projectors in the same context, are listed in Table~\ref{tab:prob-init}. The recorded probabilities give: 
\begin{align}
    \sum_{k=1}^4 P(1|k) = 0.9906, \quad \sum_{k=5}^8 P(1|k) = 0.9861, \quad \sum_{k=9}^{12} P(1|k) = 0.9816, \quad \sum_{k=13}^{16} P(1|k) = 0.9768.
\end{align}
We see that the first three probabilities agree with the three constraints in Eq.\,(3) well, and the fourth probability strongly favors the prediction of quantum mechanics over the noncontextuality theories. If we look at the sum of all the probabilities, we still have $\sum_{k=1}^{16}P(1|k)=3.9351$ exceeding the classical budget given by taking the sum of the probabilities in Eq.\,(3) and assume the noncontextuality prediction; in contrast, the total probability only falls short of the quantum prediction by 0.0649 due to experimental imperfections like dark counts and finite extinction ratio of projective measurements.

Our purpose of observing the Hardy-type paradox here was to establish the picture of logical contextuality and intuitively demonstrate the inconsistency between the predictions of noncontextuality and quantum theories. A quantified test of the paradox would rely on the condition that all the relations of orthogonality, as specified by the graph of exclusivity, strictly hold, which was never the case experimentally. For the same reason, we did not analyze the statistical significance of the observed Hardy-type paradox; the high-dimensional contextuality will be quantified using the contextuality witness in the next section.

\begin{table}[htbp]
    \centering
    \caption{Recorded detection probabilities of the initial state $\ket{\psi}= (\ket{1}+\ket{2}+\ket{3}+\ket{4})/2$ on various projectors defined in Table~\ref{tab:settings}. These probabilities establish the Hardy-type argument in Eq.\,(3) in the main text---the sum of all the probabilities markedly exceeded the budget allowed by any classical (noncontextuality) theory and was close to the prediction of quantum theory. The statistical significance of the data in the ``subtotal'' and ``total'' columns are not analyzed because the projectors in each context did not have ideal exclusivity; that is, the observed Hardy-type contextuality was qualitative. \\} 
    \begin{tabular}{cccccccccc}
    \toprule Context & \!Term\! & Measured result & \!Term\! & Measured result & \!Term\! & Measured result & \!Term\! & Measured result & Subtotal \\
    \midrule
    1 & $P(1|1)$ & 0.2384 $\!\pm\!$ 0.0014 & $P(1|2)$ & 0.2534 $\!\pm\!$ 0.0014 & $P(1|3)$ & 0.2415 $\!\pm\!$ 0.0014 & $P(1|4)$ & 0.2574 $\!\pm\!$ 0.0014 & 0.9906 \\
    2 & $P(1|5)$ & 0.2585 $\!\pm\!$ 0.0015 & $P(1|6)$ & 0.2410 $\!\pm\!$ 0.0014 & $P(1|7)$ & 0.2512 $\!\pm\!$ 0.0014 & $P(1|8)$ & 0.2353 $\!\pm\!$ 0.0014 & 0.9861 \\
    3 & $P(1|9)$ & 0.2439 $\!\pm\!$ 0.0014 & $P(1|{10})$ & 0.2536 $\!\pm\!$ 0.0014 & $P(1|{11})$ & 0.2431 $\!\pm\!$ 0.0014 & $P(1|{12})$ & 0.241 $\!\pm\!$ 0.0014 & 0.9816 \\
    4 & $P(1|{13})$ & 0.2337 $\!\pm\!$ 0.0014 & $P(1|{14})$ & 0.2437 $\!\pm\!$ 0.0014 & $P(1|{15})$ & 0.2538 $\!\pm\!$ 0.0014 & $P(1|{16})$ & 0.2456 $\!\pm\!$ 0.0014 & 0.9768 \\
    \bottomrule
    \multicolumn{9}{r}{Total} & 3.9351 \\
    \multicolumn{9}{r}{Classical budget} & 3 \\
    \multicolumn{9}{r}{Quantum prediction} & 4 \\
    \cmidrule(lr){8-10}
    \end{tabular}
    \label{tab:prob-init}
\end{table}

\paragraph{Calculation of the contextuality witness $\mu_3$.---}We used the noncontextuality inequality Eq.\,(4) in the main text to quantify the degree of contextuality in our experiment. The terms on the left-hand side of the inequality only consist of two types of probabilities: (i) $\{P(1|k)\mid k\in V(G_3)\}$, the detection probabilities of the initial state on the $+1$-eigenstate of the 16 projectors that already appeared in Table~\ref{tab:prob-init}, and (ii) $\{P(1,1|i,j) \mid (i,j)\in E(G_3)\})$, the probabilities of two ideally exclusive projective measurements both give result 1 when implemented sequentially. Experimentally, these probabilities were determined by the decomposition of sequential measurements into demolition measurements and the repreparation of post-measurement states, which gives:
\begin{align}
    P(1,1|i,j) = P(1|i)\,P(1|i=1,j),
\end{align}
where $P(1|i=1,j)$ indicates the detection probability of the $+1$-eigenstate of $\Pi_i=\ket{v_i}\bra{v_i}$ (that is, the post-measurement state when the outcome of $\Pi_i$ is 1) on the $+1$-eigenstate of $\Pi_j=\ket{v_j}\bra{v_j}$. By this decomposition, we were able to calculate all the probabilities in Eq.\,(4) in the main text using the recorded probabilities $P(1|i)$ as given in Table \ref{tab:prob-init} and $P(1|i=1,j)$ as given in Table \ref{tab:prob-vert}. Our results gave $\mu_3 \geqslant 3.821 \pm 0.012$ violating the prediction of noncontextuality by 68.7 standard deviations. Comparing the ratio between the experimental result and the noncontextuality bound, we found a ratio of $\mu_3/\alpha \geqslant 1.273$. To our best knowledge, this is the largest quantum--classical ratio ever reported in a single-particle test of contextuality; see Section \ref{sec:comparison} for a corroboration.

\begin{table}[htbp]
    \centering
    \caption{Joint detection probabilities of exclusive projectors. The probability $P(1|i=1, j)$ indicates the detection probability of the $+1$-eigenstate of $\Pi_i=\ket{v_i}\bra{v_i}$ (that is, the post-measurement state when the outcome of $\Pi_i$ is 1) on the $+1$-eigenstate of $\Pi_j=\ket{v_j}\bra{v_j}$. We would have $\Pi_i \Pi_j = 0$ across any row of the table if the conditions of orthogonality, given by the graph of exclusivity, are strict. The non-zero values thus indicate deviations from ideal exclusivity and were exploited to bound the value of the contextuality witness $\mu_3$ using the inequality \,(4) in the main text. \\}
    \begin{tabular}{cccc|cccc}
    \toprule 
    Index & Edge $(i,j)$ & $P(1|i=1,j)~(\%)$ & $P(1|j=1,i)~(\%)$ & Index & Edge $(i,j)$ & $P(1|i=1,j)~(\%)$ & $P(1|j=1,i)~(\%)$ \\
    \midrule
    1 & $(1, 2)$ & 1.211\,$\pm$\,0.031 & 1.079\,$\pm$\,0.030 & 37 & $(5, 16)$ & 0.995\,$\pm$\,0.028 & 1.150\,$\pm$\,0.030 \\
    2 & $(1, 3)$ & 0.053\,$\pm$\,0.007 & 0.093\,$\pm$\,0.009 & 38 & $(6, 7)$ & 0.151\,$\pm$\,0.011 & 0.268\,$\pm$\,0.015 \\
    3 & $(1, 4)$ & 0.036\,$\pm$\,0.005 & 0.062\,$\pm$\,0.007 & 39 & $(6, 8)$ & 0.239\,$\pm$\,0.014 & 0.420\,$\pm$\,0.018 \\
    4 & $(1, 5)$ & 0.161\,$\pm$\,0.011 & 0.067\,$\pm$\,0.007 & 40 & $(6, 10)$ & 1.154\,$\pm$\,0.031 & 1.131\,$\pm$\,0.030 \\
    5 & $(1, 6)$ & 0.126\,$\pm$\,0.010 & 0.040\,$\pm$\,0.006 & 41 & $(6, 11)$ & 0.902\,$\pm$\,0.027 & 0.971\,$\pm$\,0.028 \\
    6 & $(1, 10)$ & 1.261\,$\pm$\,0.032 & 1.674\,$\pm$\,0.037 & 42 & $(6, 13)$ & 1.163\,$\pm$\,0.031 & 1.323\,$\pm$\,0.033 \\
    7 & $(1, 12)$ & 0.264\,$\pm$\,0.015 & 0.411\,$\pm$\,0.018 & 43 & $(6, 15)$ & 1.031\,$\pm$\,0.029 & 0.995\,$\pm$\,0.028 \\
    8 & $(1, 14)$ & 0.229\,$\pm$\,0.014 & 0.158\,$\pm$\,0.011 & 44 & $(7, 8)$ & 0.344\,$\pm$\,0.017 & 0.260\,$\pm$\,0.015 \\
    9 & $(1, 15)$ & 1.536\,$\pm$\,0.035 & 1.188\,$\pm$\,0.031 & 45 & $(7, 10)$ & 0.951\,$\pm$\,0.028 & 0.639\,$\pm$\,0.023 \\
    10 & $(2, 3)$ & 0.699\,$\pm$\,0.024 & 0.807\,$\pm$\,0.026 & 46 & $(7, 11)$ & 0.597\,$\pm$\,0.022 & 0.512\,$\pm$\,0.020 \\
    11 & $(2, 4)$ & 0.040\,$\pm$\,0.006 & 0.071\,$\pm$\,0.008 & 47 & $(7, 14)$ & 0.283\,$\pm$\,0.015 & 0.387\,$\pm$\,0.018 \\
    12 & $(2, 5)$ & 0.203\,$\pm$\,0.013 & 0.250\,$\pm$\,0.014 & 48 & $(7, 16)$ & 0.587\,$\pm$\,0.022 & 0.568\,$\pm$\,0.021 \\
    13 & $(2, 6)$ & 0.469\,$\pm$\,0.019 & 0.441\,$\pm$\,0.019 & 49 & $(8, 9)$ & 0.330\,$\pm$\,0.016 & 0.464\,$\pm$\,0.019 \\
    14 & $(2, 9)$ & 0.318\,$\pm$\,0.016 & 0.493\,$\pm$\,0.020 & 50 & $(8, 12)$ & 0.269\,$\pm$\,0.015 & 0.285\,$\pm$\,0.015 \\
    15 & $(2, 11)$ & 0.384\,$\pm$\,0.018 & 0.442\,$\pm$\,0.019 & 51 & $(8, 13)$ & 0.648\,$\pm$\,0.023 & 0.619\,$\pm$\,0.022 \\
    16 & $(2, 13)$ & 0.619\,$\pm$\,0.022 & 0.493\,$\pm$\,0.020 & 52 & $(8, 15)$ & 0.558\,$\pm$\,0.021 & 0.634\,$\pm$\,0.023 \\
    17 & $(2, 16)$ & 0.815\,$\pm$\,0.026 & 0.388\,$\pm$\,0.018 & 53 & $(9, 10)$ & 1.180\,$\pm$\,0.031 & 0.991\,$\pm$\,0.028 \\
    18 & $(3, 4)$ & 1.475\,$\pm$\,0.035 & 1.526\,$\pm$\,0.035 & 54 & $(9, 11)$ & 0.665\,$\pm$\,0.023 & 0.608\,$\pm$\,0.022 \\
    19 & $(3, 7)$ & 0.222\,$\pm$\,0.013 & 0.375\,$\pm$\,0.017 & 55 & $(9, 12)$ & 0.340\,$\pm$\,0.017 & 0.209\,$\pm$\,0.013 \\
    20 & $(3, 8)$ & 0.375\,$\pm$\,0.017 & 0.378\,$\pm$\,0.017 & 56 & $(9, 13)$ & 0.507\,$\pm$\,0.020 & 0.509\,$\pm$\,0.020 \\
    21 & $(3, 10)$ & 0.767\,$\pm$\,0.025 & 0.803\,$\pm$\,0.025 & 57 & $(9, 14)$ & 0.521\,$\pm$\,0.021 & 0.745\,$\pm$\,0.025 \\
    22 & $(3, 12)$ & 0.434\,$\pm$\,0.019 & 0.547\,$\pm$\,0.021 & 58 & $(10, 11)$ & 0.759\,$\pm$\,0.025 & 0.832\,$\pm$\,0.026 \\
    23 & $(3, 13)$ & 0.737\,$\pm$\,0.024 & 0.669\,$\pm$\,0.023 & 59 & $(10, 12)$ & 0.504\,$\pm$\,0.020 & 0.646\,$\pm$\,0.023 \\
    24 & $(3, 16)$ & 0.878\,$\pm$\,0.027 & 0.513\,$\pm$\,0.020 & 60 & $(10, 13)$ & 1.232\,$\pm$\,0.032 & 0.864\,$\pm$\,0.026 \\
    25 & $(4, 7)$ & 0.509\,$\pm$\,0.020 & 0.400\,$\pm$\,0.018 & 61 & $(10, 14)$ & 0.406\,$\pm$\,0.018 & 0.326\,$\pm$\,0.016 \\
    26 & $(4, 8)$ & 0.568\,$\pm$\,0.021 & 0.374\,$\pm$\,0.017 & 62 & $(11, 12)$ & 0.423\,$\pm$\,0.018 & 0.291\,$\pm$\,0.015 \\
    27 & $(4, 9)$ & 1.074\,$\pm$\,0.029 & 1.064\,$\pm$\,0.029 & 63 & $(11, 15)$ & 1.026\,$\pm$\,0.029 & 0.838\,$\pm$\,0.026 \\
    28 & $(4, 11)$ & 0.692\,$\pm$\,0.024 & 0.632\,$\pm$\,0.023 & 64 & $(11, 16)$ & 0.222\,$\pm$\,0.013 & 0.183\,$\pm$\,0.012 \\
    29 & $(4, 14)$ & 0.673\,$\pm$\,0.023 & 0.680\,$\pm$\,0.023 & 65 & $(12, 15)$ & 0.392\,$\pm$\,0.018 & 0.456\,$\pm$\,0.019 \\
    30 & $(4, 15)$ & 0.981\,$\pm$\,0.028 & 0.941\,$\pm$\,0.028 & 66 & $(12, 16)$ & 0.508\,$\pm$\,0.020 & 0.643\,$\pm$\,0.023 \\
    31 & $(5, 6)$ & 1.699\,$\pm$\,0.037 & 2.147\,$\pm$\,0.042 & 67 & $(13, 14)$ & 0.440\,$\pm$\,0.019 & 0.395\,$\pm$\,0.018 \\
    32 & $(5, 7)$ & 0.082\,$\pm$\,0.008 & 0.221\,$\pm$\,0.013 & 68 & $(13, 15)$ & 1.192\,$\pm$\,0.031 & 1.057\,$\pm$\,0.029 \\
    33 & $(5, 8)$ & 0.293\,$\pm$\,0.015 & 0.298\,$\pm$\,0.016 & 69 & $(13, 16)$ & 0.610\,$\pm$\,0.022 & 0.494\,$\pm$\,0.020 \\
    34 & $(5, 9)$ & 1.001\,$\pm$\,0.028 & 1.041\,$\pm$\,0.029 & 70 & $(14, 15)$ & 0.706\,$\pm$\,0.024 & 0.846\,$\pm$\,0.026 \\
    35 & $(5, 12)$ & 1.487\,$\pm$\,0.035 & 1.026\,$\pm$\,0.029 & 71 & $(14, 16)$ & 0.043\,$\pm$\,0.006 & 0.158\,$\pm$\,0.011 \\
    36 & $(5, 14)$ & 0.789\,$\pm$\,0.025 & 0.961\,$\pm$\,0.028 & 72 & $(15, 16)$ & 1.481\,$\pm$\,0.035 & 1.257\,$\pm$\,0.032 \\
    \bottomrule
    \end{tabular}
    \label{tab:prob-vert}
\end{table}

\begin{table}[htbp]
    \centering
    \caption{Detection probabilities of post-measurement states on compatible projectors. The probability $P(1|i=0, j)$ indicates the detection probability of the state $(\mathbb{I}-\Pi_i)\ket{\psi}/\Tr[(\mathbb{I}-\Pi_i)\ket{\psi}\bra{\psi}]$ (that is, the post-measurement state of $\ket{\psi}$ when the measurement outcome of $\Pi_i$ is 0) on the $+1$-eigenstate of $\Pi_j=\ket{v_j}\bra{v_j}$. The theoretical values are $1/3$ for all entries across the table. The quantities here, together with those in Table \ref{tab:prob-init} and \ref{tab:prob-vert}, were exploited to estimate the value of the signaling factors $\varepsilon_{ij}^\prime$ using Eq.\,(5) in the main text. \\}
    \begin{tabular}{cccc|cccc}
    \toprule 
    Index & Edge $(i,j)$ & $P(1|i=0,j)~(\%)$ & $P(1|j=0,i)~(\%)$ & Index & Edge $(i,j)$ & $P(1|i=0,j)~(\%)$ & $P(1|j=0,i)~(\%)$ \\
    \midrule
    1 & $(1, 2)$ & 32.47\,$\pm$\,1.40 & 30.37\,$\pm$\,1.37 & 37 & $(5, 16)$ & 36.04\,$\pm$\,1.39 & 38.23\,$\pm$\,1.42 \\
    2 & $(1, 3)$ & 30.88\,$\pm$\,1.39 & 28.93\,$\pm$\,1.37 & 38 & $(6, 7)$ & 34.17\,$\pm$\,1.42 & 34.08\,$\pm$\,1.42 \\
    3 & $(1, 4)$ & 35.06\,$\pm$\,1.42 & 33.46\,$\pm$\,1.40 & 39 & $(6, 8)$ & 29.36\,$\pm$\,1.38 & 32.02\,$\pm$\,1.42 \\
    4 & $(1, 5)$ & 33.67\,$\pm$\,1.40 & 31.04\,$\pm$\,1.37 & 40 & $(6, 10)$ & 33.58\,$\pm$\,1.41 & 32.79\,$\pm$\,1.39 \\
    5 & $(1, 6)$ & 29.57\,$\pm$\,1.38 & 28.86\,$\pm$\,1.37 & 41 & $(6, 11)$ & 28.35\,$\pm$\,1.36 & 31.16\,$\pm$\,1.39 \\
    6 & $(1, 10)$ & 32.13\,$\pm$\,1.39 & 29.03\,$\pm$\,1.36 & 42 & $(6, 13)$ & 29.58\,$\pm$\,1.39 & 32.16\,$\pm$\,1.42 \\
    7 & $(1, 12)$ & 31.35\,$\pm$\,1.40 & 28.97\,$\pm$\,1.37 & 43 & $(6, 15)$ & 33.14\,$\pm$\,1.40 & 31.19\,$\pm$\,1.37 \\
    8 & $(1, 14)$ & 29.46\,$\pm$\,1.37 & 29.90\,$\pm$\,1.38 & 44 & $(7, 8)$ & 33.29\,$\pm$\,1.40 & 31.12\,$\pm$\,1.37 \\
    9 & $(1, 15)$ & 32.86\,$\pm$\,1.40 & 29.03\,$\pm$\,1.36 & 45 & $(7, 10)$ & 33.56\,$\pm$\,1.37 & 34.49\,$\pm$\,1.38 \\
    10 & $(2, 3)$ & 33.85\,$\pm$\,1.38 & 36.56\,$\pm$\,1.42 & 46 & $(7, 11)$ & 29.76\,$\pm$\,1.34 & 31.67\,$\pm$\,1.36 \\
    11 & $(2, 4)$ & 36.82\,$\pm$\,1.39 & 37.22\,$\pm$\,1.40 & 47 & $(7, 14)$ & 33.38\,$\pm$\,1.38 & 31.49\,$\pm$\,1.36 \\
    12 & $(2, 5)$ & 35.23\,$\pm$\,1.37 & 35.22\,$\pm$\,1.37 & 48 & $(7, 16)$ & 34.39\,$\pm$\,1.39 & 33.73\,$\pm$\,1.38 \\
    13 & $(2, 6)$ & 33.87\,$\pm$\,1.38 & 33.21\,$\pm$\,1.38 & 49 & $(8, 9)$ & 29.19\,$\pm$\,1.39 & 28.98\,$\pm$\,1.38 \\
    14 & $(2, 9)$ & 35.04\,$\pm$\,1.39 & 37.06\,$\pm$\,1.42 & 50 & $(8, 12)$ & 28.06\,$\pm$\,1.38 & 30.26\,$\pm$\,1.40 \\
    15 & $(2, 11)$ & 31.43\,$\pm$\,1.35 & 32.64\,$\pm$\,1.36 & 51 & $(8, 13)$ & 29.34\,$\pm$\,1.40 & 30.49\,$\pm$\,1.42 \\
    16 & $(2, 13)$ & 32.26\,$\pm$\,1.38 & 34.89\,$\pm$\,1.41 & 52 & $(8, 15)$ & 30.44\,$\pm$\,1.39 & 28.45\,$\pm$\,1.36 \\
    17 & $(2, 16)$ & 34.17\,$\pm$\,1.38 & 35.96\,$\pm$\,1.40 & 53 & $(9, 10)$ & 34.78\,$\pm$\,1.41 & 31.76\,$\pm$\,1.37 \\
    18 & $(3, 4)$ & 34.59\,$\pm$\,1.41 & 33.14\,$\pm$\,1.39 & 54 & $(9, 11)$ & 31.94\,$\pm$\,1.39 & 31.44\,$\pm$\,1.38 \\
    19 & $(3, 7)$ & 33.21\,$\pm$\,1.40 & 31.54\,$\pm$\,1.38 & 55 & $(9, 12)$ & 31.48\,$\pm$\,1.38 & 33.42\,$\pm$\,1.41 \\
    20 & $(3, 8)$ & 30.43\,$\pm$\,1.39 & 32.82\,$\pm$\,1.43 & 56 & $(9, 13)$ & 30.45\,$\pm$\,1.38 & 31.68\,$\pm$\,1.40 \\
    21 & $(3, 10)$ & 34.54\,$\pm$\,1.41 & 32.38\,$\pm$\,1.38 & 57 & $(9, 14)$ & 32.84\,$\pm$\,1.40 & 29.55\,$\pm$\,1.36 \\
    22 & $(3, 12)$ & 31.69\,$\pm$\,1.40 & 29.22\,$\pm$\,1.36 & 58 & $(10, 11)$ & 34.71\,$\pm$\,1.39 & 34.23\,$\pm$\,1.38 \\
    23 & $(3, 13)$ & 32.00\,$\pm$\,1.42 & 29.65\,$\pm$\,1.39 & 59 & $(10, 12)$ & 31.41\,$\pm$\,1.35 & 33.64\,$\pm$\,1.38 \\
    24 & $(3, 16)$ & 32.16\,$\pm$\,1.40 & 32.66\,$\pm$\,1.40 & 60 & $(10, 13)$ & 33.27\,$\pm$\,1.39 & 34.62\,$\pm$\,1.41 \\
    25 & $(4, 7)$ & 33.84\,$\pm$\,1.35 & 36.50\,$\pm$\,1.39 & 61 & $(10, 14)$ & 34.43\,$\pm$\,1.38 & 33.58\,$\pm$\,1.38 \\
    26 & $(4, 8)$ & 34.56\,$\pm$\,1.39 & 37.33\,$\pm$\,1.43 & 62 & $(11, 12)$ & 27.55\,$\pm$\,1.34 & 31.30\,$\pm$\,1.38 \\
    27 & $(4, 9)$ & 32.97\,$\pm$\,1.36 & 33.60\,$\pm$\,1.36 & 63 & $(11, 15)$ & 32.06\,$\pm$\,1.37 & 33.66\,$\pm$\,1.40 \\
    28 & $(4, 11)$ & 32.83\,$\pm$\,1.36 & 34.60\,$\pm$\,1.38 & 64 & $(11, 16)$ & 34.52\,$\pm$\,1.42 & 29.60\,$\pm$\,1.36 \\
    29 & $(4, 14)$ & 33.83\,$\pm$\,1.37 & 34.59\,$\pm$\,1.38 & 65 & $(12, 15)$ & 31.82\,$\pm$\,1.38 & 32.87\,$\pm$\,1.39 \\
    30 & $(4, 15)$ & 34.87\,$\pm$\,1.36 & 36.80\,$\pm$\,1.39 & 66 & $(12, 16)$ & 34.48\,$\pm$\,1.43 & 29.03\,$\pm$\,1.36 \\
    31 & $(5, 6)$ & 34.34\,$\pm$\,1.38 & 36.60\,$\pm$\,1.41 & 67 & $(13, 14)$ & 28.33\,$\pm$\,1.38 & 29.03\,$\pm$\,1.39 \\
    32 & $(5, 7)$ & 36.72\,$\pm$\,1.39 & 36.13\,$\pm$\,1.38 & 68 & $(13, 15)$ & 31.16\,$\pm$\,1.40 & 25.77\,$\pm$\,1.33 \\
    33 & $(5, 8)$ & 35.34\,$\pm$\,1.40 & 36.37\,$\pm$\,1.41 & 69 & $(13, 16)$ & 33.96\,$\pm$\,1.45 & 29.38\,$\pm$\,1.39 \\
    34 & $(5, 9)$ & 35.61\,$\pm$\,1.39 & 36.78\,$\pm$\,1.40 & 70 & $(14, 15)$ & 32.32\,$\pm$\,1.38 & 33.59\,$\pm$\,1.39 \\
    35 & $(5, 12)$ & 34.19\,$\pm$\,1.38 & 35.25\,$\pm$\,1.38 & 71 & $(14, 16)$ & 32.90\,$\pm$\,1.40 & 29.57\,$\pm$\,1.36 \\
    36 & $(5, 14)$ & 34.53\,$\pm$\,1.38 & 33.37\,$\pm$\,1.36 & 72 & $(15, 16)$ & 33.69\,$\pm$\,1.38 & 34.55\,$\pm$\,1.39 \\
    \bottomrule
    \end{tabular}
    \label{tab:prob-luders}
\end{table}

\bigskip

\paragraph{Verification of the no-signaling condition.---}Although the nullification test of the signaling factors does not affect the values of the contextuality witness, it is nonetheless crucial for revealing the failure of the noncontextuality models and understanding the experiment: doing a test of contextuality means we are testing against the noncontextual hidden-variable models, in which observables have predefined context-insensitive values. If the measurement of a preceding observable disturbs the probability distribution of the succeeding observable, it can no longer be considered revealing the predefined values of the observables, and the observed phenomenon will not be able to serve as a test of the noncontextuality theory. By implementing the no-signaling test, we showed our measurements in the experiment fulfill the assumptions of no-disturbance in the noncontextuality theory; thus, it is sensible to be considered a test of the theory itself.

To determine the signaling factors, we rewrite the probabilities in Eq.\,(5) of the main text in terms of prepare-and-measure probabilities using measurement decompositions:
\begin{alignat}{3}
    \varepsilon_{ij} &= P(1|i) - P(1,1|i,j) - P(1,0|i,j) &&= P(1|i) - P(1|i) [P(1|j) + P(0|j)] = 0, \\
    \varepsilon_{ij}^\prime &= P(1|j) - P(1,1|i,j) - P(0,1|i,j) &&= P(1|j) - P(1|i)P(1|i=1, j) - [1 - P(1|i)] P(1|i=0, j).
    \label{eq:signal-decomp}
\end{alignat}
Therefore, the factors $\varepsilon_{ij}$ signifying the effect of a measurement on its preceding measurement always evaluate to zero. It is expected since no measurement has the effect of signaling backward in time. On the other hand, the factors $\varepsilon_{ij}^\prime$ signifying the effect of a measurement on its succeeding measurement are not guaranteed to cancel out, so the vanish of these factors is nontrivial and indicates no-disturbance between two measurements and the measurements have plausible compatibility.

Calculation of the quantities in Eq.\,(\ref{eq:signal-decomp}) only requires one more type of probability that is not specified in the previous text. It is $P(1|i=0, j)$, the detection probability of the state $(\mathbb{I}-\Pi_i)\ket{\psi}/\Tr[(\mathbb{I}-\Pi_i)\ket{\psi}\bra{\psi}]$ (that is, the post-measurement state of $\ket{\psi}$ when the measurement outcome of $\Pi_i$ is 0) on the $+1$-eigenstate of $\Pi_j=\ket{v_j}\bra{v_j}$. The experimental results of $P(1|i=0, j)$ are given in Table~\ref{tab:prob-luders} which, with the other probabilities already given previously, are used to calculate the signaling factors $\varepsilon_{ij}^\prime$ in the Fig.~4 of the main text. Our results specifically confirm that the destructive measurement and repreparation procedure adopted here, modulo some unavoidable experimental imperfections, does not affect the probability distribution of observables. Therefore, even though we did not use the sequential measurements experimentally, we were still able to assume the L\"uders rule, which is itself noncontextual, as the method for repreparation of the post-measurement state and test quantum contextuality.

\section{Summary of results: single-particle tests of contextuality}
\label{sec:comparison}

\begin{figure}[htbp]
    \centering
    \includegraphics[width = .99 \textwidth]{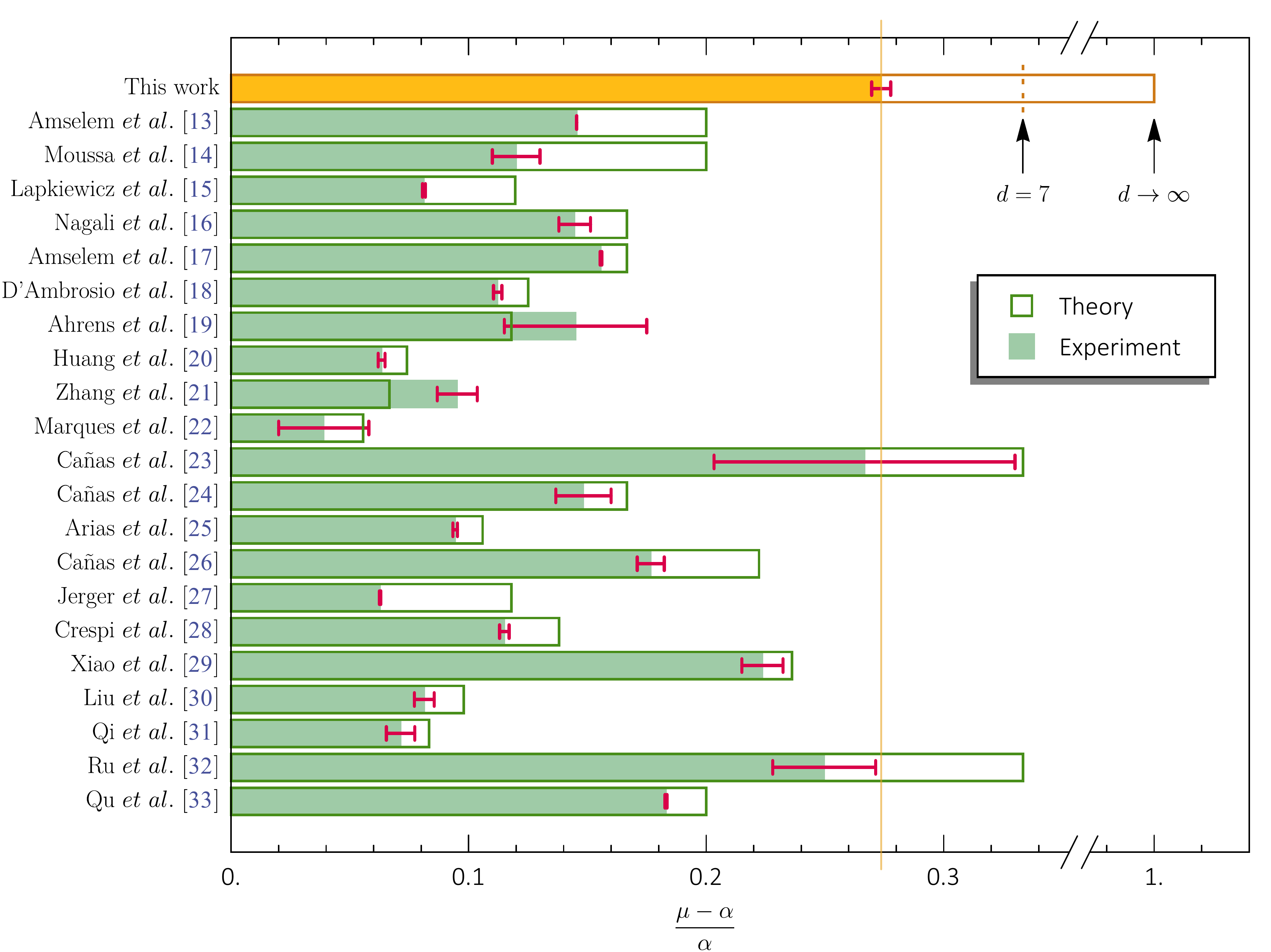}
    \caption{Comparison of the results in single-particle tests of contextuality, in the present work and from literature~\cite{s-Amselem09, s-Moussa10, s-Lapkiewicz11, s-Nagali12, s-Amselem12, s-dAmbrosio13, s-Ahrens13, s-Huang13, s-Zhang13, s-Marques14, s-Canas14, s-Canas15, s-Arias15, s-Canas16, s-Jerger16, s-Crespi17, s-Xiao18, s-Liu19, s-Qi22, s-Ru22, s-Qu21}. The degree of contextuality is quantified by the violation of the noncontextuality inequality in the Cabello--Severini--Winter form normalized by the classical bound. Here, $\mu$ indicates the observed left-hand side value of the noncontextuality inequality, and $\alpha$ is the independence number of the graph of exclusivity corresponding to the tested noncontextuality (and thus its classical bound). The error bars represent the $1\sigma$ standard deviation of the experimental results. The frame and filling of the bars give the theoretical maximum and experimental results. For the present work, the solid and dashed theoretical bounds correspond to an infinite-dimensional system and a 7-dimensional system, respectively.}
    \label{fig:compare}
\end{figure}

In Fig.~\ref{fig:compare}, we compare the degree of contextuality observed in our experiment with the results of other works reported in the literature~\cite{s-Amselem09, s-Moussa10, s-Lapkiewicz11, s-Nagali12, s-Amselem12, s-dAmbrosio13, s-Ahrens13, s-Huang13, s-Zhang13, s-Marques14, s-Canas14, s-Canas15, s-Arias15, s-Canas16, s-Jerger16, s-Crespi17, s-Xiao18, s-Liu19, s-Qi22, s-Ru22, s-Qu21}. The degree of contextuality is quantified by the violation ratio of the noncontextuality inequality in the Cabello--Severini--Winter (sum of event probability) form~\cite{s-CSW14}. The ratio has the physical implication of how much error every individual projective measurement can tolerate before the phenomenon of contextuality disappears; in this way, the results in different experiments can be compared in a consistent sense. 

It is evident that the results in different works are presented in different ways, and to make the comparison we need to convert the results into the measurement probability form. Here, we shall use the experimental results reported in Ref.~\cite{s-Lapkiewicz11} to illustrate how the conversion works.

The noncontextuality inequality tested in Ref.~\cite{s-Lapkiewicz11} reads:
\begin{align}
    \braket{A_1 A_2} + \braket{A_2 A_3} + \braket{A_3 A_4} + \braket{A_4 A_5} + \braket{A_5 A_1^\prime} - \braket{A_1 A_1^\prime} \overset{\rm NCHV}{\geqslant} -4,
    \label{eq:kcbs6-corr}
\end{align}
where the $d$-dimensional observables $A_1, \ldots, A_5$ and $A_1^\prime$ have one eigenvalue being $-1$ and the other $d-1$ eigenvalues being $+1$. The equation was derived by introducing a correction term to the Klyachko--Can--Binicio\u{g}lu--Shumovsky noncontextuality inequality to account for the different realizations of the same measurement in different contexts. The correlations are related to the probabilities of element events by $\braket{A_j A_k} = P(A_i=+1, A_j=+1) + P(A_i=-1, A_j=-1) - P(A_i=-1, A_j=+1) - P(A_i=+1, A_j=-1).$ If we only keep the probabilities of element events which \textit{increases} the violation of the inequality, we can exploit the completeness of the probabilities to write:
\begin{align}
    \braket{A_j A_k} = 1 - 2\,[P(A_i=-1, A_j=+1) - P(A_i=+1, A_j=-1)],
    \label{eq:corr2prob}
\end{align}
and substituting the above relation into Eq.\,(\ref{eq:kcbs6-corr}), which yields:
\begin{align}
    \sum_{k=1}^4 P(A_k=+1,A_{k+1}=-1) + \sum_{k=1}^4 P(A_k=-1,A_{k+1}=+1) + P(A_1=+1, A_1^\prime=+1) + P(A_1=-1, A_1^\prime=-1) \overset{\rm NCHV}{\leqslant} 5.
    \label{eq:kcbs6-prob}
\end{align}
The inequality is but a noncontextuality inequality in the measurement probability form; for this kind of inequality, the noncontextuality bound of the quantum maximum can be determined using the graph-theoretic approach to quantum correlations. 

\begin{figure}[htbp]
    \centering
    \includegraphics[width = .3 \textwidth]{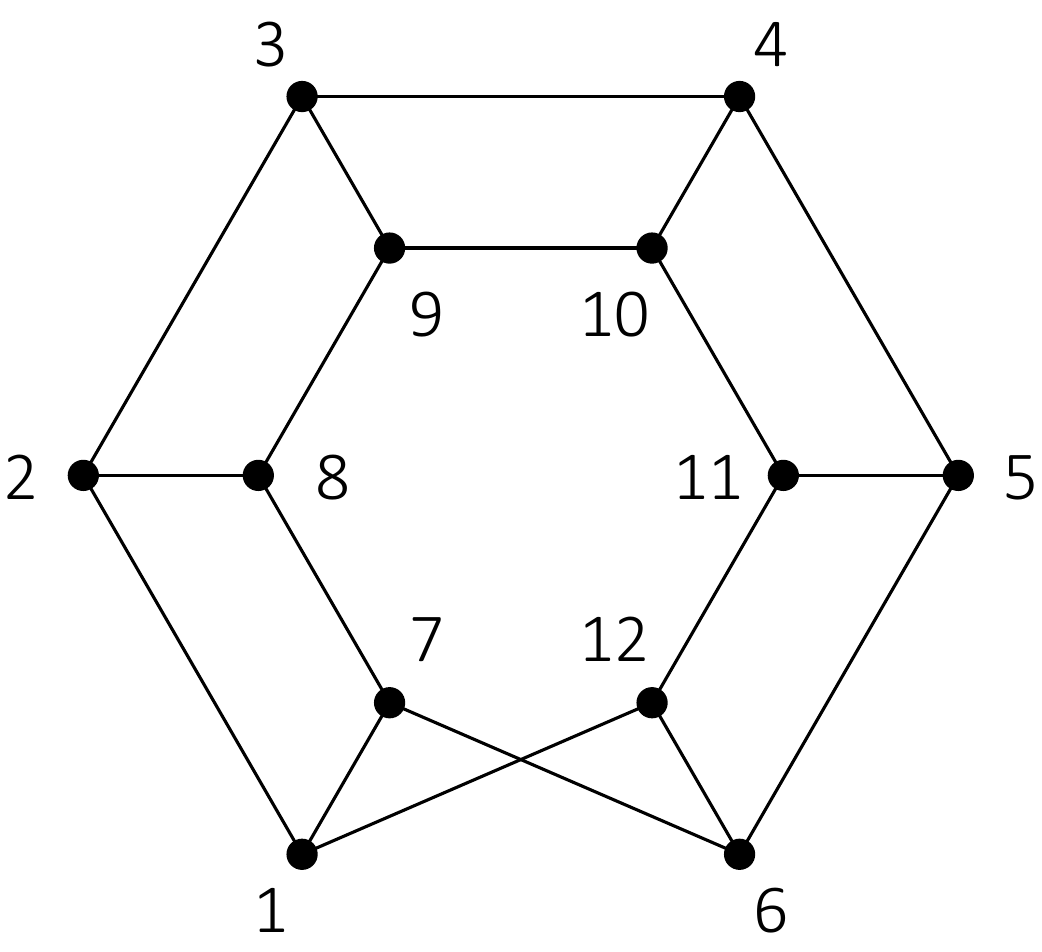}
    \caption{The graph of exclusivity of the events appeared in Ref.~\cite{s-Lapkiewicz11}. The definition of the events are: 
    1: $A_1=+1, A_2=-1$, 2: $A_2=+1, A_3=-1$, 3:  $A_3=+1, A_4=-1$, 4: $A_4=+1, A_5=-1$, 5: $A_5=+1, A_1^\prime=-1$, 6: $A_1^\prime=+1, A_1=+1$, 
    7: $A_1=-1, A_2=+1$, 8: $A_2=-1, A_3=+1$, 9:  $A_3=-1, A_4=+1$, 10: $A_4=-1, A_5=+1$, 11: $A_5=-1, A_1^\prime=+1$, 12: $A_1^\prime=-1, A_1=-1$.} 
    \label{fig:moebius6}
\end{figure}

We plot the graph of exclusivity here in Fig.~\ref{fig:moebius6}. It is an M\"obius ladder $M_{12}$ with a length of 6. The independence number of the graph is $\alpha(M_{12}) = 5$---it is also the classical bound in Eq.\,(\ref{eq:kcbs6-prob}). The quantum maximum of Eq.\,(\ref{eq:kcbs6-prob}) is the Lov\'asz number of the graph of exclusivity, which was recently proved for the M\"obius ladders to be~\cite{s-Bharti22} $$\vartheta(M_{2n}) = \frac{n}{2}\left(1+\cos\frac{\pi}{n}\right); $$
substituting $n=6$ into the expression we have $\vartheta(M_{12})=3+3\sqrt{3}/2 \approx 5.598$. On the experiment side, we use the inverse transformation of Eq.\,(\ref{eq:corr2prob}) to calculate the experimental left-hand side value of the inequality (\ref{eq:kcbs6-prob}) from the data already available in Ref.~\cite{s-Lapkiewicz11}. We found the reported result achieved a violation ratio of 8.12\% against the noncontextuality models.

Using the methods above, we have evaluated the violation of noncontextuality from the other works in the sum of event probability picture. In particular, we calculate the left-hand side value, $\mu$, of the noncontextuality inequality in each of the works using the reported results, and use the graph of exclusivity of the measurements to determine the classical bound $\alpha$ and the quantum maximum $\vartheta$. The plotted quantities in Fig.~\ref{fig:compare} are the experimental ratio of violation of the noncontextuality inequality, $(\mu-\alpha)/\alpha$ and the maximal ratio of violation violation allowed by the quantum theory, $(\vartheta-\alpha)/\alpha$. We note that in Ref.~\cite{s-Amselem09, s-Moussa10, s-Lapkiewicz11} and \cite{s-Ahrens13, s-Zhang13}, the reported data were not enough to determine the orthogonality of ideally exclusive measurements, and we did not subtract the effect therein; this causes the latter two works appear to have an observed correlation stronger than the quantum maximum. In Ref.~\cite{s-Canas15, s-Ru22}, the theoretical bound of violation is the same as the case of $d=7$ in our work; however, our experiment used a lower Hilbert space dimension than both of the two previous works, and our construction grows stronger when the Hilbert space dimension further increases. Overall, the comparison here corroborated our previous affirmation that the result reported here is the highest degree of contextuality ever observed on a single system.


\begin{thebibliography}{99}
\bibitem{note2}
Graph states are a family of highly entangled multiqubit states represented by undirected graphs, not to be confused with the graph of exclusivity in the graph-theoretic approach of contextuality.
\end{thebibliography}
\end{document}